\pdfoutput=1

\documentclass[conference]{IEEEtran}
\IEEEoverridecommandlockouts
\usepackage{cite}
\usepackage{amsmath,amssymb,amsfonts}

\usepackage{multirow}
\usepackage{enumitem}
\usepackage[hyphens]{url}
\usepackage{fancyhdr}
\usepackage{colortbl}
\usepackage{hyperref}
\usepackage{wrapfig}
\usepackage{algorithm2e}
\RestyleAlgo{ruled}
\usepackage{pifont}
\usepackage{makecell}
\usepackage{threeparttable}
\usepackage{tikz}
\usepackage{float}
\usepackage{booktabs}
\usepackage{cleveref}
\usepackage{comment}
\usepackage{subcaption}

\newcommand*\circled[1]{\tikz[baseline=(char.base)]{
            \node[shape=circle,fill,inner sep=0.2pt] (char) {\textcolor{white}{#1}};}}

\newcommand{\NAME}{SCALE-Sim }
\newcommand{\NAMEnospace}{SCALE-Sim}

\crefformat{section}{\S#2#1#3} 
\crefformat{subsection}{\S#2#1#3}
\crefformat{subsubsection}{\S#2#1#3}
\def\BibTeX{{\rm B\kern-.05em{\sc i\kern-.025em b}\kern-.08em
    T\kern-.1667em\lower.7ex\hbox{E}\kern-.125emX}}
\begin{document}

\title{\NAME v3: A modular cycle-accurate systolic accelerator simulator for end-to-end system analysis}

\author{
    \IEEEauthorblockN{Ritik Raj\IEEEauthorrefmark{1}\thanks{\IEEEauthorrefmark{4}The authors contributed equally to this work.}, 
    Sarbartha Banerjee\IEEEauthorrefmark{2}\IEEEauthorrefmark{4}, 
    Nikhil Chandra\IEEEauthorrefmark{1}\IEEEauthorrefmark{4}, 
    Zishen Wan\IEEEauthorrefmark{1}\IEEEauthorrefmark{4}, 
    Jianming Tong\IEEEauthorrefmark{1}\IEEEauthorrefmark{4}, \\
    Ananda Samajdar\IEEEauthorrefmark{3},
    Tushar Krishna\IEEEauthorrefmark{1}}
    \IEEEauthorblockA{\IEEEauthorrefmark{1}Georgia Institute of Technology}
    \IEEEauthorblockA{\IEEEauthorrefmark{2}University of Texas Austin}
    \IEEEauthorblockA{\IEEEauthorrefmark{3}IBM Research}
}

\maketitle
\thispagestyle{plain}
\begin{abstract}
The rapid advancements in AI, scientific computing, and high-performance computing (HPC) have driven the need for versatile and efficient hardware accelerators. Existing tools like \NAME v2 provide valuable cycle-accurate simulations for systolic-array-based architectures but fall short in supporting key modern features such as sparsity, multi-core scalability, and comprehensive memory analysis. To address these limitations, we present \NAME v3 (\href{https://github.com/scalesim-project/scale-sim-v3}{GitHub Repository}), a modular, cycle-accurate simulator that extends the capabilities of its predecessor. \NAME v3 introduces five significant enhancements: multi-core simulation with spatio-temporal partitioning and hierarchical memory structures, support for sparse matrix multiplications (SpMM) with layer-wise and row-wise sparsity, integration with Ramulator for detailed DRAM analysis, precise data layout modeling to minimize memory stalls, and energy and power estimation via Accelergy. These improvements enable deeper end-to-end system analysis for modern AI accelerators, accommodating a wide variety of systems and workloads and providing detailed full-system insights into latency, bandwidth, and power efficiency.

A 128×128 array is 6.53× faster than a 32×32 array for ViT-base, using only latency as a metric. However, \NAME v3 finds that 32×32 is 2.86× more energy-efficient due to better utilization and lower leakage energy. For EdP, 64×64 outperforms both 128×128 and 32×32 for ViT-base. \NAME v2 shows a 21\% reduction in compute cycles for six ResNet18 layers using weight-stationary (WS) dataflow compared to output-stationary (OS). However, when factoring in DRAM stalls, OS dataflow exhibits 30.1\% lower execution cycles compared to WS, highlighting the critical role of detailed DRAM analysis.

\end{abstract}



\section{Introduction}


The computation capacity has been increasing significantly over the years to drive algorithm advances in the field of Artificial Intelligence, Scientific computing, High-Performance Computing (HPC), AR/VR applications, and so on. As per ~\cite{sevilla2022compute}, the training compute demands (FLOPS) have been doubling every nine months since 2015. There has been a plethora of accelerators targeting matrix multiplication in AI \cite{jouppi2023tpu, Cerebras, firoozshahian2023mtia, lichtenau2022ai, reuther2020survey}, Scientific computing \cite{feinberg2018enabling, sun2022burstz+, bocco2019smurf, weber2010comparing}, HPC \cite{ujaldon2016hpc, temam2012defect, britt2017quantum, li2022vector}, AR/VR \cite{yang2022three, sumbul2023fully, li2024fusion}, and robotic \cite{krishnan2022automatic, hao2024orianna, liu2022energy, liu2021robotic} applications. A fixed configuration for such accelerators is inefficient for all the varieties of workloads. These differences have led to an era of domain-specific AI accelerators, where it is necessary for accelerators to adapt to their specific workloads.


Designing an efficient AI accelerator is a challenging problem that requires a deeper analysis of computing hardware, domain-specific workloads, target constraints such as latency, power, and bandwidth as well as mapping strategies. To mitigate this problem, there exist several tools \cite{parashar2019timeloop, munoz2021stonne, samajdar2018scale, kwon2020maestro, wu2022usystolic} for modeling performance and energy behavior of AI accelerators. \NAME (SystoliC AcceLErator SIMulator) v2 \cite{samajdar2018scale} is one of the popular open-source simulators for systolic array-based accelerators in use today. Fundamentally, it models the runtime for convolution and GEMM operators - which are the key building blocks within modern AI models.
It also provides the flexibility to support systolic arrays of different shapes and sizes in addition to the sizes of double-buffered on-chip memories for input activations, weights, and output activations. In addition, it supports flexibility in mapping operators to the systolic array via three classic dataflows~\cite{chen2016eyeriss} - input stationary, weight stationary, and output stationary. \NAME v2 provides a detailed output report consisting of latency, bandwidth requirement, and cycle-accurate read/write traces for SRAM and main memory (DRAM or HBM).


\begin{figure*}
    \centering
    \includegraphics[width=\linewidth]{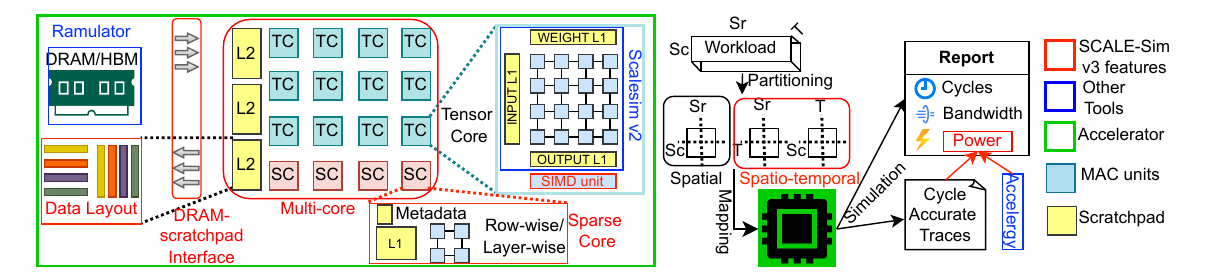}
    \caption{Overview of \NAME v3 highlighting the new features over \NAME v2 marked in red}
    \label{fig:scalesim v3 overview}
\end{figure*}

\begin{table*}[]
    \centering
    \begin{threeparttable}
    \resizebox{\textwidth}{!}{
    \begin{tabular}{|c|c|c|c|c|c|c|c|c|}
    \hline
        Framework & Compute & Cores & Partitioning & Sparse  & Power & Energy Model & Data & Main Memory \\
        & & & & Core & & & Layout & Model \\
\hline

MAESTRO \cite{kwon2019understanding, kwon2020maestro} & analytical & single & N/A & N/A & N/A & Cacti-6.0 \cite{muralimanohar2009cacti} & No & N/A \\
\hline
STONNE \cite{munoz2021stonne} & cycle-accurate & single & N/A & SIGMA- & Average & Table-based model & No & N/A \\
& & & & like \cite{qin2020sigma} & & based on synthesis &  & \\
& & & & Unstructured & & and place-and-route  &  &\\

\hline
        Timeloop v4 \cite{parashar2019timeloop} & analytical & many & Spatio-temporal & Distribution & Average & Accelergy \cite{wu2019accelergy} & No \tnote{*} & CactiDRAM \cite{balasubramonian2017cacti}
        \\
        & & & & based & & &  &\\

\hline
        \NAME v2 \cite{samajdar2018scale} & cycle-accurate & many & Spatial & N/A & N/A & N/A & No & N/A
        \\
\hline
        \textbf{\NAME v3} & cycle-accurate & many & Spatio-temporal & Layer and Row & Instantaneous + & Accelergy \cite{wu2019accelergy} & yes & Ramulator \cite{kim2015ramulator} \\
        & & & & wise N:M & Average & & & \\
\hline
    \end{tabular}}
    \caption{Related Works}
        \label{tab:related works}

    \begin{tablenotes}
      \footnotesize
      \item[*] FEATHER~\cite{tong2024featherreconfigurableacceleratordata} paper adds data layout modeling support in Timeloop but is not part of Timeloop's public repo yet.
    \end{tablenotes}
      \end{threeparttable}
\vspace{-2em}
\end{table*}

\NAME v2 is a promising tool for exploring the design space of systolic-array-based AI accelerators and has been leveraged by
several recent works \cite{kim2022ark, ma2023camj, qin2020sigma, guo2023cambricon,samajdar2022self}. However, in this work, we identify limitations in \NAME v2 that restrict it from modeling recent innovations in AI accelerators and enabling full-system analysis and comparison of diverse designs, as we discuss next.

\textbf{Limitation 1: Multiple Tensor Cores.} Given a mix of both matrix and vector operations in modern AI workloads, recent accelerators leverage a mix of matrix multiplication engines (i.e., systolic arrays) and vector engines. Furthermore, to support more flexible partitioning and mapping, recent accelerators employ \textit{multiple} tensor cores backed by a shared scratchpad memory. 
Examples include Google's TPUv4~\cite{jouppi2023tpu} with two tensor cores, Meta's MTIA~\cite{firoozshahian2023mtia} with 64 PEs (each PE has two RISC-V cores)  Cerebras WSE-2~\cite{Cerebras} with 850k sparse tensor cores.
\NAME v2 does not have support to model such chips, in part because when it was proposed that most accelerators just had single systolic arrays. 

\textbf{Limitation 2: Sparsity.} In recent years, sparsity has become a key feature in state-of-the-art AI models \cite{liu2023deja, gale2019state, liu2015sparse}, enabling a suite of software optimizations \cite{gale2023megablocks, shen2023efficient}, hardware accelerators \cite{giannoula2022towards, ampere,liu2015sparse, jeong2023vegeta, ampere} as well as hardware-software co-design \cite{kanellopoulos2019smash, zhou2018cambricon} techniques to enhance computational efficiency.
Furthermore, there is also an increasing interest in leveraging sparse matrix multiplication accelerators for domains beyond AI - such as scientific computing and graph analytics which are inherently hypersparse.
Leveraging sparsity can reduce unnecessary computations and memory accesses, thereby enhancing computational efficiency and energy consumption. 
Unfortunately, \NAME v2 does not support sparse matrix multiplication units, which are now present in modern AI accelerators like TPU v5 \cite{tpuv5}.

\textbf{Limitation 3: Main Memory Interface.}  The main memory technology implementation~\cite{jun2017hbm} and the workload's specific access pattern can lead to considerable variations in memory read/write latency, energy, and row conflicts. In fact, this behavior is exploited by attacks such as row hammer \cite{mutlu2017rowhammer, mutlu2019rowhammer}. 
Therefore, a deeper study of the DRAM/HBM and on-chip memory interface is needed to understand the memory behavior of different workloads running on an AI accelerator. 
Unfortunately, \NAME v2 models main memory as a monolithic entity with a fixed latency and bandwidth.
The lack of a main memory interface in \NAME v2 limits it from providing the key statistics mentioned above. 

\textbf{Limitation 4: Data Layout.} The organization of data within on-chip storage, such as SRAM or BRAM, plays a critical role in the performance of machine learning accelerators. The interaction between data layout and dataflow can significantly impact latency and compute efficiency~\cite{tong2024featherreconfigurableacceleratordata}, a ignorance of the actual data layout could lead to a magnitude of performance off. Optimal dataflows, which define how data is processed and moved through compute and memory resources, can be hindered by suboptimal data layouts due to issues like bank conflicts—when multiple data elements are accessed from the same memory bank simultaneously, causing stalls in computation. Therefore, a detailed modeling of data layout on the memory behavior is essential which is overlooked by \NAME v2. 

\textbf{Limitation 5. Energy.} 
In recent years, hardware accelerators have become a key approach for improving energy efficiency by leveraging the unique characteristics of specific application domains. However, accurately predicting energy consumption often requires completing the physical design layout, which significantly hinders design space exploration due to high overhead and slow simulation times~\cite{wu2019accelergy}. To enable faster and more efficient design exploration, energy estimation must be performed earlier in the design process, without requiring a fully detailed hardware description.
Additionally, data movement energy constitutes a significant portion of an accelerator's total energy consumption, particularly in data-dependent scenarios. Unfortunately, \NAME v2 lacks support for energy and power modeling, limiting its ability to provide valuable insights for designing energy-efficient accelerators.

In this work, we address the aforementioned limitations of \NAME v2 and enhance it via five new features as shown in \autoref{fig:scalesim v3 overview}. \circled{1} Modeling of multi-core features including spatio-temporal partitioning, L2 shared memory, heterogeneous tensor cores and non-uniform workload partitioning. \circled{2} Support for systolic-array based sparse accelerators targeting layer-wise and row-wise SpMM workloads. \circled{3} DRAM and on-chip memory interface modeled by Ramulator \cite{kim2015ramulator} integration. \circled{4} Modeling of data layout for a more accurate on-chip memory stall analysis. \circled{5} Energy and power analysis modeled by Accelergy \cite{wu2019accelergy} integration.

\section{Background}

\subsection{Systolic arrays and multi-core accelerators}
Systolic arrays are specialized hardware architectures designed to efficiently execute repetitive computations, particularly those involving linear algebra operations such as matrix multiplication and convolutions \cite{kung1979systolic, kung1982systolic, mead1980introduction, saptalakar2013design, chiper2002systolic, quinton1983systematic, johnson1993general} commonly found in digital signal processing and machine learning workloads. 


Research indicates that computational requirements for AI models have doubled every 3-4 months since 2012 \cite{openai_compute}. These requirements are driven by the development of increasingly complex models and the expansion of AI applications across various sectors including natural language processing (NLP) \cite{dubey2024llama, team2023gemini, achiam2023gpt}, autonomous driving \cite{zheng2025genad, hu2023planning, atakishiyev2024explainable}, healthcare \cite{rahman2023ambiguous, zhang2022contrastive, gao2018human} and so on. This exponential growth necessitates specialized hardware accelerators. A lot of AI accelerators, including Google TPU v5 \cite{tpuv5}, Meta MTIA \cite{firoozshahian2023mtia}, Cerebras WSE-2 \cite{Cerebras}, Nvidia DGX GH200 \cite{nvidia-dgx}, and Tesla Dojo \cite{talpes2022dojo}  have incorporated multi-chip and multi-core designs to meet the increasing demands. 

\vspace{-0.5em}





\subsection{Sparse Accelerators}




Sparsity has been widely adopted in commercial accelerators due to its potential to enhance computational efficiency. For example, NVIDIA's Ampere architecture \cite{ampere} introduced Sparse Tensor Cores, which leverage structured sparsity with a 2:4 sparsity ratio, ensuring that two out of every four tensor elements are non-zero. Similarly, Google TPUs \cite{tpuv5} utilize sparse accelerators to efficiently process sparse tensors, incorporating mechanisms like sparse weight pruning and sparse matrix operations to scale effectively for large AI workloads. Accelerators such as VerSA\cite{seo2024versa}  and VEGETA~\cite{jeong2023vegeta} propose a versatile systolic array capable of accelerating both dense and sparse matrix multiplications, addressing the diverse needs of modern deep neural network applications.
\vspace{-0.5em}



\subsection{Memory Simulators}
Memory simulators like Ramulator~\cite{ramulatorv2,kim2015ramulator} or DRAMSim3~\cite{dramsim3} simulate the internal behavior of DDR and other memory technologies. 
Typically these simulators provide detailed metrics such as total memory requests, read/write operations, average latency, bandwidth utilization, row buffer hit rate, and power consumption estimates, aiding in comprehensive memory system performance analysis.
Specifically, we integrate Ramulator~\cite{kim2015ramulator} as the memory model for \NAME v3, which is a high-performance, cycle-accurate DRAM simulator facilitating the evaluation and design of memory systems.
Its modular architecture allows for straightforward integration of various DRAM standards, including DDR3, DDR4, LPDDR4, GDDR5, WIO1, WIO2, and HBM, enabling researchers to assess and compare different memory technologies within a unified framework. 
\vspace{-0.5em}

\subsection{Accelergy}
Accelergy \cite{wu2019accelergy} is an architecture-level energy estimation tool designed to provide accurate and flexible energy consumption analyses for accelerator designs. Developed to address the challenges of energy-efficient computing, Accelergy allows designers to model both primitive components and complex, user-defined compound components within an accelerator architecture. It systematically captures the energy consumption of various building blocks, facilitating rapid design space exploration without the need for detailed physical layouts, achieving up to 95\% accuracy in estimating energy consumption for well-known deep neural network accelerators like Eyeriss \cite{chen2016eyeriss}. 
\vspace{-0.5em}


\subsection{\NAME v2}

\NAME v2 is a cycle-accurate simulator designed to model systolic array-based accelerators for convolutions and GEMMs. The simulator accepts user-defined configurations, allowing for the specification of parameters including systolic array dimensions, on-chip double-buffered SRAM sizes, dataflow depending on reuse (input, weight or output stationary), and DRAM bandwidth. It processes neural network topologies provided in CSV format, detailing layer-wise specifications to accurately model workloads. Upon execution, \NAME v2 generates detailed reports, including compute cycles, bandwidth utilization, and cycle-accurate SRAM and DRAM traces, offering insights into the performance characteristics of the simulated architecture.


In the rest of the paper, we present the five enhancements we add to \NAME v2 - namely support for multiple tensor cores, sparsity, main memory, data layouts, and energy/power.
\section{Multiple Tensor Cores}
\label{sec:mulicore}
This section describes the multi tensor core simulation feature of \NAME v3 which is orthogonal to v2 in the following four ways: \circled{1} \NAME v3 expands the spatial workload partitioning mentioned in v2 to spatio-temporal partitioning (\autoref{subsection:spatio-temporal}) as shown in \autoref{fig:partitioning}. \circled{2} The memory on the chip is modeled in a hierarchical structure where multiple cores share an L2 scratchpad (\autoref{subsection:hierarchial memory}), as shown in \autoref{fig:shared_L2}. \circled{3} Heterogeneous tensor cores (\autoref{subsection:heterogenous tensor}) in terms of systolic array dimensions and SIMD length. \circled{4} Non-uniform workload partitioning (\autoref{subsection:non uniform partitioning}) for cores having different latency profiles \cite{shao2019simba}

\subsection{Spatio-temporal partitioning}
\label{subsection:spatio-temporal}
\vspace{-1.5em}
\begin{figure}[H]
    \centering
\begin{table}[H]
\centering
\resizebox{0.6\columnwidth}{!}{%
\begin{tabular}{|l|lll|}
\hline
\multicolumn{1}{|c|}{\multirow{2}{*}{Dataflow}} & \multicolumn{3}{c|}{Mapping}                          \\ \cline{2-4} 
\multicolumn{1}{|c|}{}                          & \multicolumn{1}{l|}{Sr} & \multicolumn{1}{l|}{Sc} & T \\ \hline
Input Stationary                                & \multicolumn{1}{l|}{K}  & \multicolumn{1}{l|}{N}  & M \\ \hline
Weight Stationary                               & \multicolumn{1}{l|}{K}  & \multicolumn{1}{l|}{M}  & N \\ \hline
Output Stationary                               & \multicolumn{1}{l|}{M}  & \multicolumn{1}{l|}{N}  & K \\ \hline
\end{tabular}%
}
\caption{GEMM mapping for different dataflows}
\label{tab:gemm}
\end{table}   
    \vspace{-1em}
     \includegraphics[width=\linewidth]{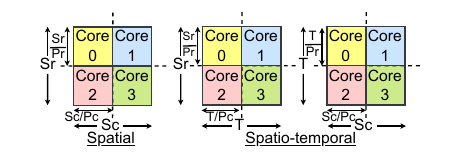}
    \caption{Spatial and spatio-temporal partitioning}
    \label{fig:partitioning}
        \vspace{-1em}
\end{figure}

We introduce spatio-temporal partitioning as a new feature, enhancing spatial partitioning introduced in the \NAME v2 paper~\cite{samajdar2018scale} as shown in \autoref{fig:partitioning}. It refers to partitioning along one of spatial (Sr/Sc) and temporal (T) dimensions (\autoref{tab:gemm}) instead of just spatial dimensions.
Assuming R and C as systolic array dimensions; Sr, Sc, and T as spatial and temporal mapping dimensions (\autoref{tab:gemm}); Pr and Pc as the no. of row and column partitions (\autoref{fig:partitioning}) such as no. of cores = Pr$\times$Pc, the following equations lists the runtime for different partitioning: 
    \vspace{-1em}

\begin{equation}
    (2*R+C+T-2) * \lceil \frac{Sr}{\frac{Pr}{R}} \rceil * \lceil \frac{Sc}{\frac{Pc}{C}} \rceil \text{\cite{samajdar2018scale}}
    \label{eq: spatial}
    \vspace{-1em}
\end{equation}

\begin{equation}
    (2*R+C+ \lceil \frac{T}{Pc} \rceil -2) * \lceil \frac{Sr}{\frac{Pr}{R}} \rceil * \lceil \frac{Sc}{Pc} \rceil
    \label{eq: st1}
        \vspace{-0.5em}
\end{equation}

\begin{equation}
    (2*R+C+ \lceil \frac{T}{Pr} \rceil -2) * \lceil \frac{Sr}{R} \rceil * \lceil \frac{Sc}{\frac{Pc}{C}} \rceil
    \label{eq: st2}
        \vspace{-0.5em}
\end{equation}

where \autoref{eq: spatial} represents spatial \cite{samajdar2018scale} while \autoref{eq: st1} and \autoref{eq: st2} represents spatio-temporal partitioning.

\begin{figure}[H]
    \centering
    \begin{minipage}{0.45\textwidth}
        \centering
        \includegraphics[width=\textwidth]{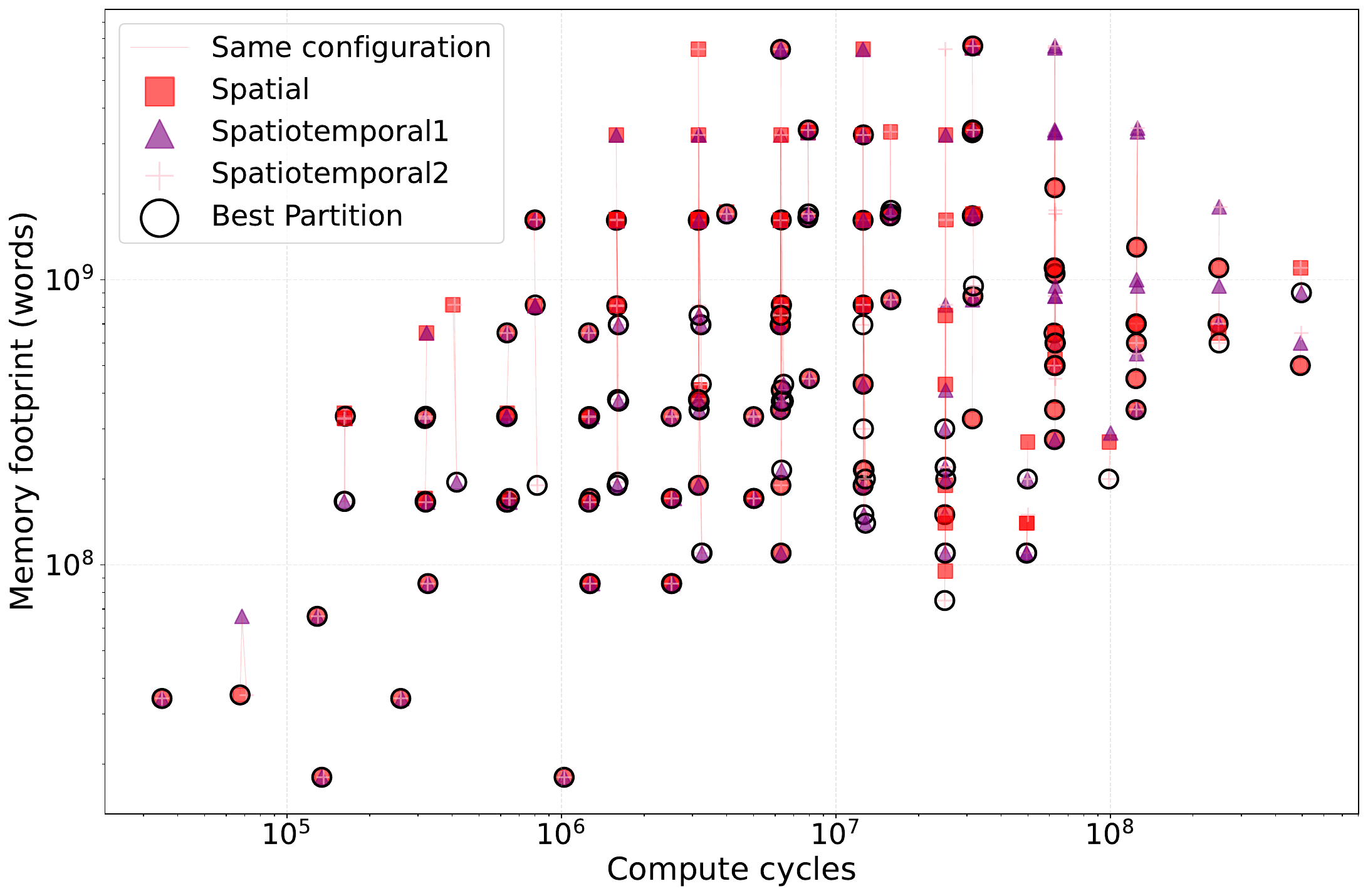}
         \\(a) Compute cycles optimized partitions (Pr, Pc)
        \label{fig:compute_optimized_spatiotemporal}
    \end{minipage}
    \hfill
    \begin{minipage}{0.45\textwidth}
        \centering
        \includegraphics[width=\textwidth]{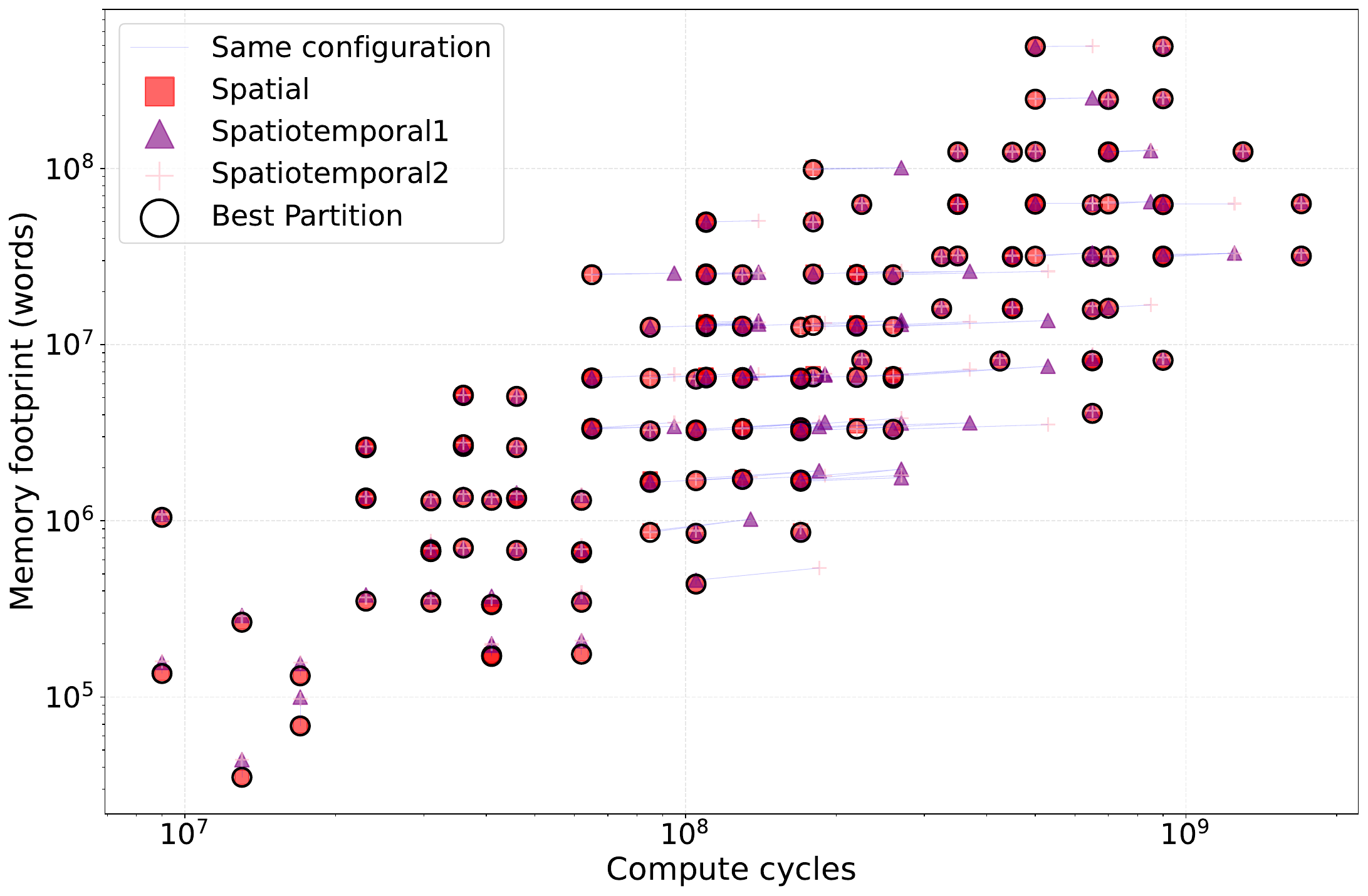}
         \\(b) Memory footprint optimized partitions (Pr, Pc)
        \label{fig:memory_optimized_spatiotemporal}
    \end{minipage}
    \caption{Compute cycles v/s memory footprint tradeoff for spatial and spatiotemporal partitioning for scale-out cases}
    \label{fig:spatiotemporal}
    \vspace{-1em}
\end{figure}

\autoref{fig:spatiotemporal} shows the memory footprint and compute cycles trade-off between spatial, spatio-temporal 1 (\autoref{eq: st1}), and spatio-temporal 2 (\autoref{eq: st2}) partitioning schemes. GEMM dimensions M, N, and K are chosen from [1000, 5000, 10000], so a total of 27 workloads. Systolic array dimensions (row and column sizes) are selected from [8, 16, 32], while the number of scale-out cores is chosen from [16, 32, 64]. Pr and Pc are chosen based on optimizing compute cycles (\hyperref[fig:spatiotemporal]{Figure 3a}) and memory footprint (\hyperref[fig:spatiotemporal]{Figure 3b}). The line connections denote that the points have the same configuration of workload, array size, and no. of cores. In \hyperref[fig:spatiotemporal]{Figure 3a} (compute-optimized), the best partition among the three connected points is the one with the least memory footprint. In \hyperref[fig:spatiotemporal]{Figure 3b} (memory-optimized), the best partition is the one with the least compute cycles. There are multiple examples in \hyperref[fig:spatiotemporal]{Figure 3a} where spatiotemporal partitioning schemes outperform spatial partitioning. However, in \hyperref[fig:spatiotemporal]{Figure 3b}, spatial partitioning outperforms in most cases. 



\subsection{Hierarchical memory with shared L2}
\label{subsection:hierarchial memory}

\begin{figure}[H]
    \centering
    \includegraphics[width=0.7\linewidth, height=0.4\linewidth]{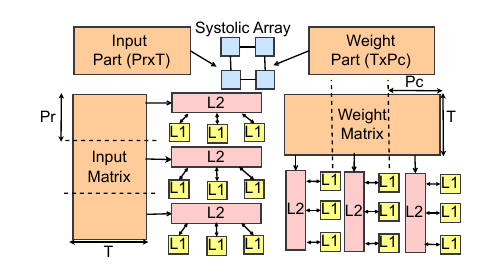}
    \caption{Shared Input and Weight L2 SRAMs}
    \label{fig:shared_L2}
\vspace{-1em}
\end{figure}

Due to spatial partitioning, each core works on input partition, Pr $\times$ T and weight partition, Pc $\times$ T as shown in \autoref{fig:shared_L2}. Each core in the same row receives the same partition of the input matrix while each column receives the same partition of the weight matrix. If there is only L1 SRAM, there will be lots of duplication across multiple cores in the same row (duplication of input matrix) or the same column (duplication of weight matrix). To mitigate the data duplication, we use shared L2 SRAM as shown in \autoref{fig:shared_L2}. To ensure no stalls, the size of L2 SRAM should be enough to accommodate the input/weight partitions. L1 SRAM size can be configured to get a good balance between L1 size and L1 misses. Similarly, the shared L2 will receive input/weight partitions with common dimension Sr and Sc in case of spatiotemporal partitioning.

\subsection{Heterogeneous Tensor Cores}
\label{subsection:heterogenous tensor}

We add the support for tensor cores, following the naming convention from Google's TPU \cite{jouppi2023tpu} where each TensorCore consists of one or more matrix-multiply units (MXUs) and a vector unit. The vector/SIMD unit is used for general computation such as activations and softmax. Similarly, SIMD unit in Meta's MTIA \cite{firoozshahian2023mtia} handles quantization/de-quantization and nonlinear functions using lookup tables and floating-point units to approximate functions like exponential, sigmoid, and tanh. Therefore, \NAME supports arbitrary tensor cores with heterogeneity in terms of the length and type of SIMD units and/or systolic array dimensions as incorporated in \cite{symons2022towards, nandakumar2022accelerating, spantidi2022targeting}. In addition, the latency of SIMD units is customization as per the use case.

\subsection{Non-uniform Workload Partitioning}
\label{subsection:non uniform partitioning}

Multi-Chip-Module \cite{lau1994chip, doane2013multichip, arunkumar2017mcm, sekhar2024multi} based accelerators including Simba \cite{shao2019simba} have a non-uniform latency profile based on the location of individual cores. The cores/chiplets which are farther away from the main memory requires more no. of hops or higher Network on Package (NoP) latency for communication of input or output matrices. On the other hand, the cores/chiplets which are closer to the main memory require lower NoP latency for input or output communication. This difference in NoP latency gives rise to non-uniform workload partitioning where farther cores/chiplets receive lower amount of workload (lower Pr and Pc) while nearer cores/chiplets receive higher amount of workload (higher Pr and Pc). \NAME supports non-uniform partitioning to effectively simulate cores/chiplets having different compute requirements. 
\vspace{-0.4em}
\section{Sparsity}
\label{sec:sparsity}
\vspace{-0.4em}


Sparsity is a fundamental characteristic observed in many deep learning models, where a significant proportion of tensor elements (such as weights or activations) are zero or negligibly small. Sparsity can be formally quantified as the ratio of non-zero valued elements in a block (N) to the total number of elements in the block (M) i.e. sparsity ratio is N:M.

\vspace{-0.4em}
\subsection{Sparsity in \NAME v3}

Building on this foundation, \NAME v3 extends its predecessor by incorporating support for layer-wise and row-wise sparsity in systolic array architectures.

\subsubsection{Layer-Wise Sparsity}
\NAME v3 allows the sparsity configurations to vary across layers in a neural network. This flexibility aligns with the observation that different layers exhibit varying degrees of sparsity. For example, initial layers in CNNs often have denser connections, while deeper layers are more amenable to higher sparsity levels. 


\subsubsection{Row-Wise Sparsity}
Row-wise sparsity enforces a fixed number of nonzero elements at row granularity~\cite{jeong2023vegeta}, thus offering more finer control over the sparsity of the workload. In \NAME v3, row-wise sparsity is implemented for various \textbf{N:M sparsity ratios}, where each group of $M$ elements in a row contains exactly $N$ nonzero values. This feature captures the realistic sparsity patterns present in real-world CNN models. \NAME v3 constraints sparsity ratios to $N \leq M/2$, ensuring that sparsity remains computationally advantageous. The density increases for $N > M/2$, negating the benefits of sparsity and approaching dense configurations. 


\subsection{\NAME v3 + Sparsity Integration}



\textbf{Step 1: Architectural and Workload Input: }
In topology files, a \textit{SparsitySupport} column has been added representing sparsity ratios in the $N:M$ format. The configuration file has a new "sparsity" section, which contains the sparsity-related architectural knobs. For layer-wise sparsity, the \textit{SparsitySupport} knob is set to \textit{true} and the \textit{OptimizedMapping} knob is set to \textit{false}. For all the simulations presented in this paper, the \textit{SparseRep} is set to \textit{ellpack\_block}. For row-wise sparsity, the \textit{OptimizedMapping} knob is set to \textit{true} and \textit{BlockSize} holds the value of $M$ in the $N:M$ ratio. The number of non-zero elements ($N$) is randomized for different rows and is kept $\leq M/2$. The dataflow is set to \textit{weight-stationary} for all the sparsity simulations.

\textbf{Step 2: Generating operand and demand matrices: }
Once the sparsity architectural knobs and details are read from the topology and the configuration files, the operand matrices shall be modified to reflect the sparsity of the model. For simplicity, we assume that the first $N$ rows have non-zero elements and the remaining $N-M$ rows have zero elements. For row-wise sparsity, each row is assigned a random sparsity ratio such that $N \leq M/2$. Instead of streaming a single input element per row in the systolic array, we would need to stream a block of input elements. This is replicated in the \textit{IFMAP} demand matrix generation by fetching corresponding addresses from \textit{IFMAP SRAM}.


\textbf{Step 3: Collecting statistics: }
The new \NAME v3 simulator built outputs a \textit{SPARSE\_REPORT.csv} that contains information and metrics such as Sparsity Representation, Original Filter Storage, and New Filter Storage (this consists of the compressed filter matrix and the metadata).



\subsection{Sparsity Case Studies}
\begin{figure}[t!]
  \centering
  \includegraphics[width=\linewidth]{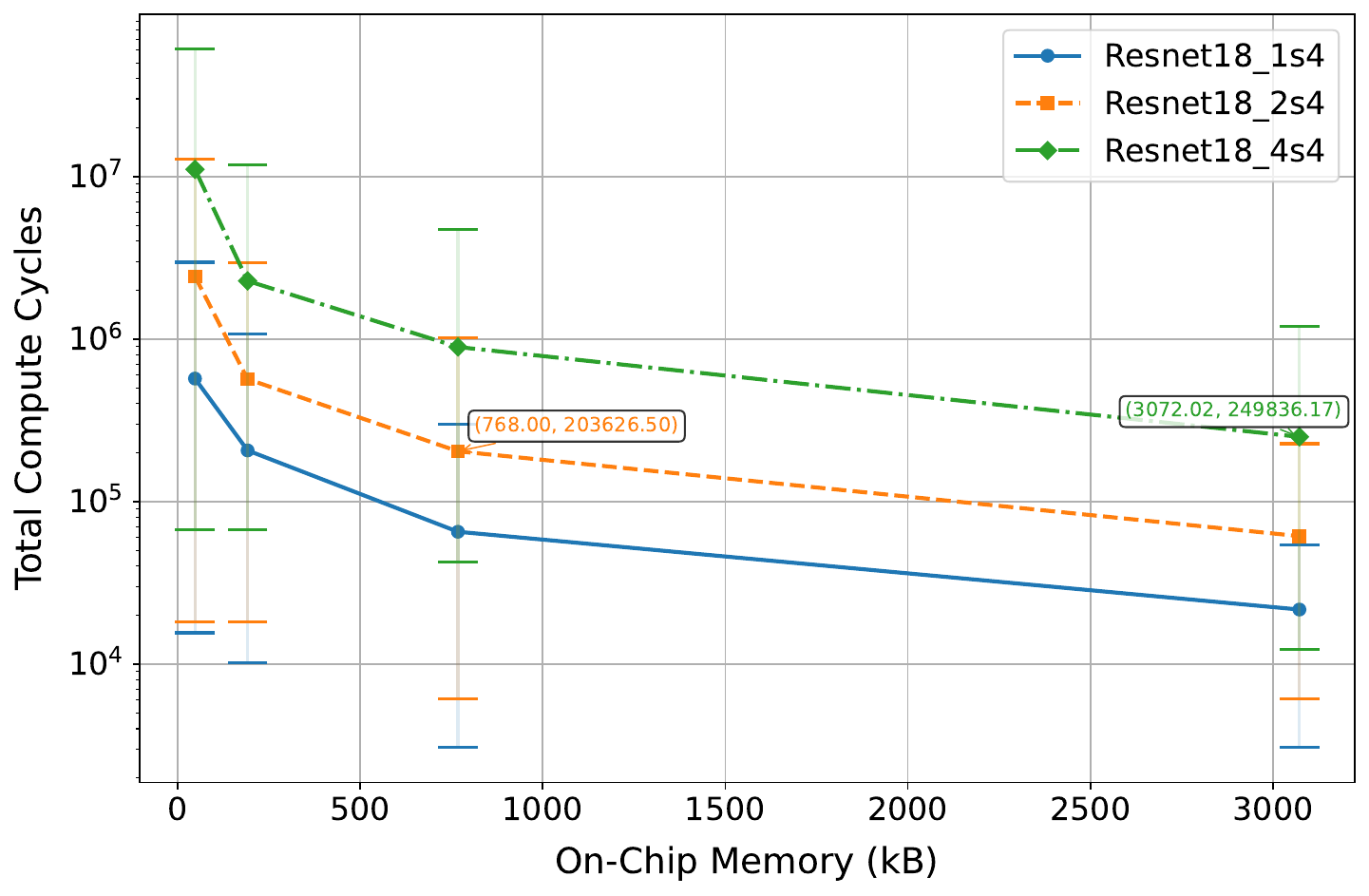}
  \caption{Total compute cycles (including memory stalls) vs On-chip memory for ResNet-18 for 1:4, 2:4 and 4:4 sparsity ratios}
  \label{fig:sparsity-lws-resnet18-computecyclesplot}
  \vspace{-0.17in}
\end{figure}

\autoref{fig:sparsity-lws-resnet18-computecyclesplot} demonstrates the relationship between total cycles including memory stalls and on-chip memory for the ResNet-18. This relationship is observed for different sparsity ratios viz., $1:4$, $2:4$ and $4:4$. As the size of on-chip memory increases, more data can be fetched into the SRAM in a single instance, which results in reduced stall cycles. For a given on-chip memory, sparse models require fewer compute cycles because of the increased sparsity.

\begin{figure}[t!]
  \centering
  \includegraphics[width=\linewidth]{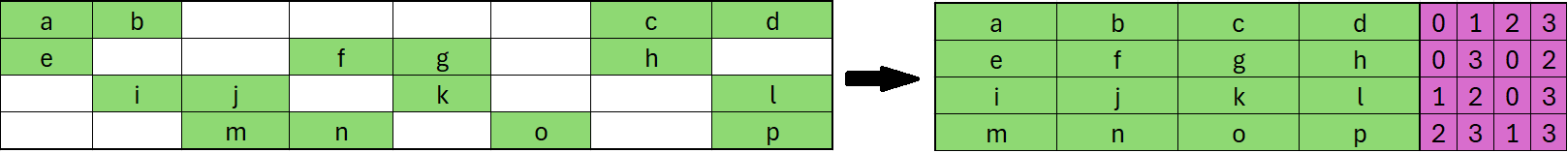}
  \caption{(a) Original matrix (b) Blocked ELLPACK format}
  \label{fig:ellpack_rep}
  \vspace{-0.17in}
\end{figure}

\NAME v3 supports Compressed Sparse Row (CSR), Compressed Sparse Column (CSC) and Blocked ELLPACK formats. For all the sparsity related simulations, Blocked ELLPACK has been considered. For a matrix shown in \hyperref[fig:ellpack_rep]{Figure 6a} and a block size equal to 4, the blocked ELLPACK representation is depicted in \hyperref[fig:ellpack_rep]{Figure 6b}; where the green cells represent the non-zero filter matrix values and the lavender cells represent the corresponding metadata. In the generic setting, the number of bits required for a single metadata entry is given by $\log_2(\text{Block Size})$.

\begin{figure}[t!]
  \centering
  \includegraphics[width=\linewidth]{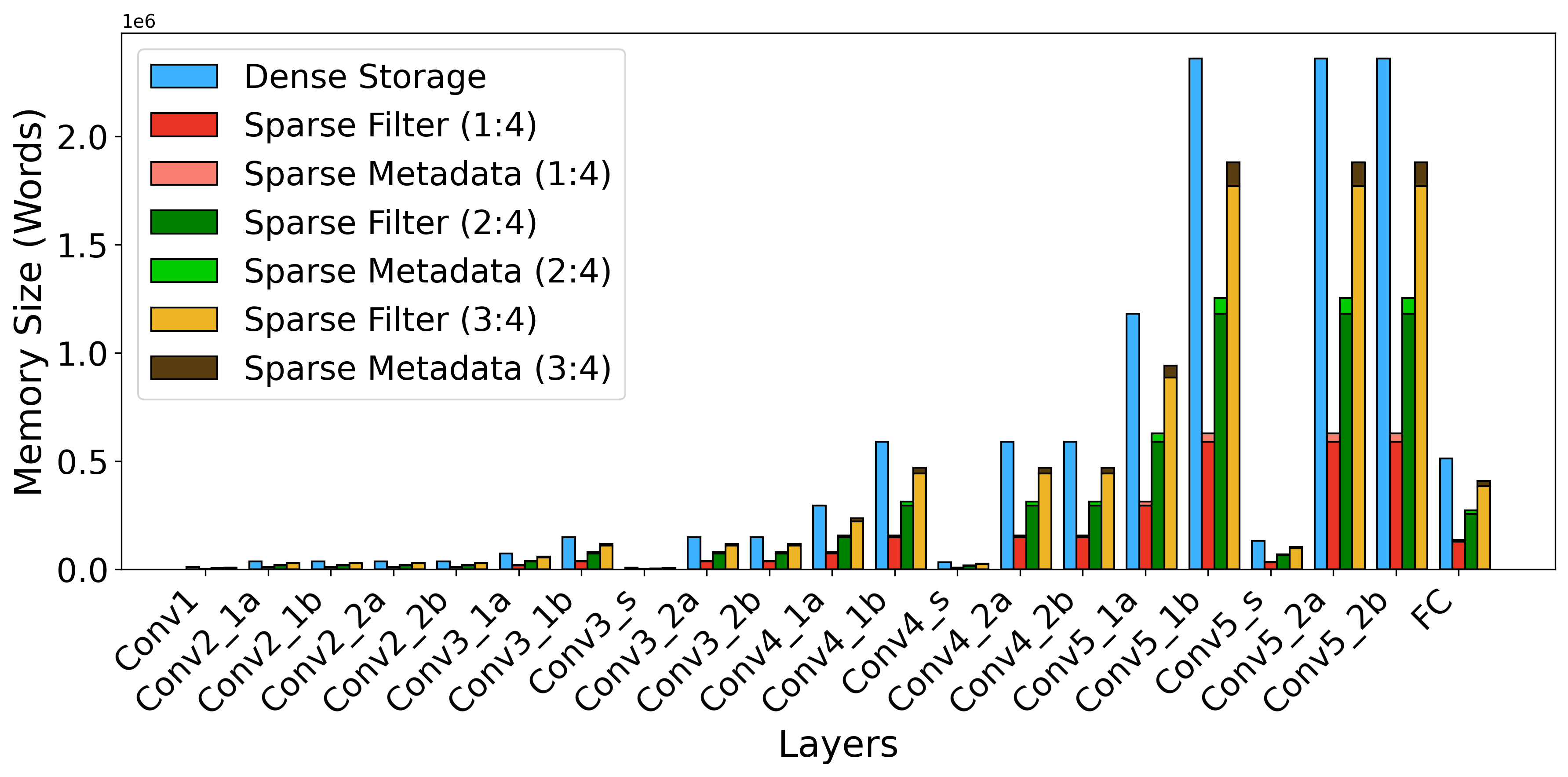}
  \caption{The memory plot shows the memory storage comparison for dense, 1:4, 2:4 and 3:4 sparsity ratios for ResNet-18}
  \label{fig:sparsity-resnet18-memoryplot}
  \vspace{-0.17in}
\end{figure}

\autoref{fig:sparsity-resnet18-memoryplot} compares the memory storage requirements of ResNet-18 layers for dense, 1:4, 2:4, and 3:4 sparsity ratios. For each sparsity ratio, the storage consists of sparse filter data and associated metadata. As sparsity increases, the number of filter elements stored in the memory and their corresponding metadata reduce highlighting the effect of sparsity on memory requirements.


\begin{figure}[t!]
  \centering
  \includegraphics[width=\linewidth]{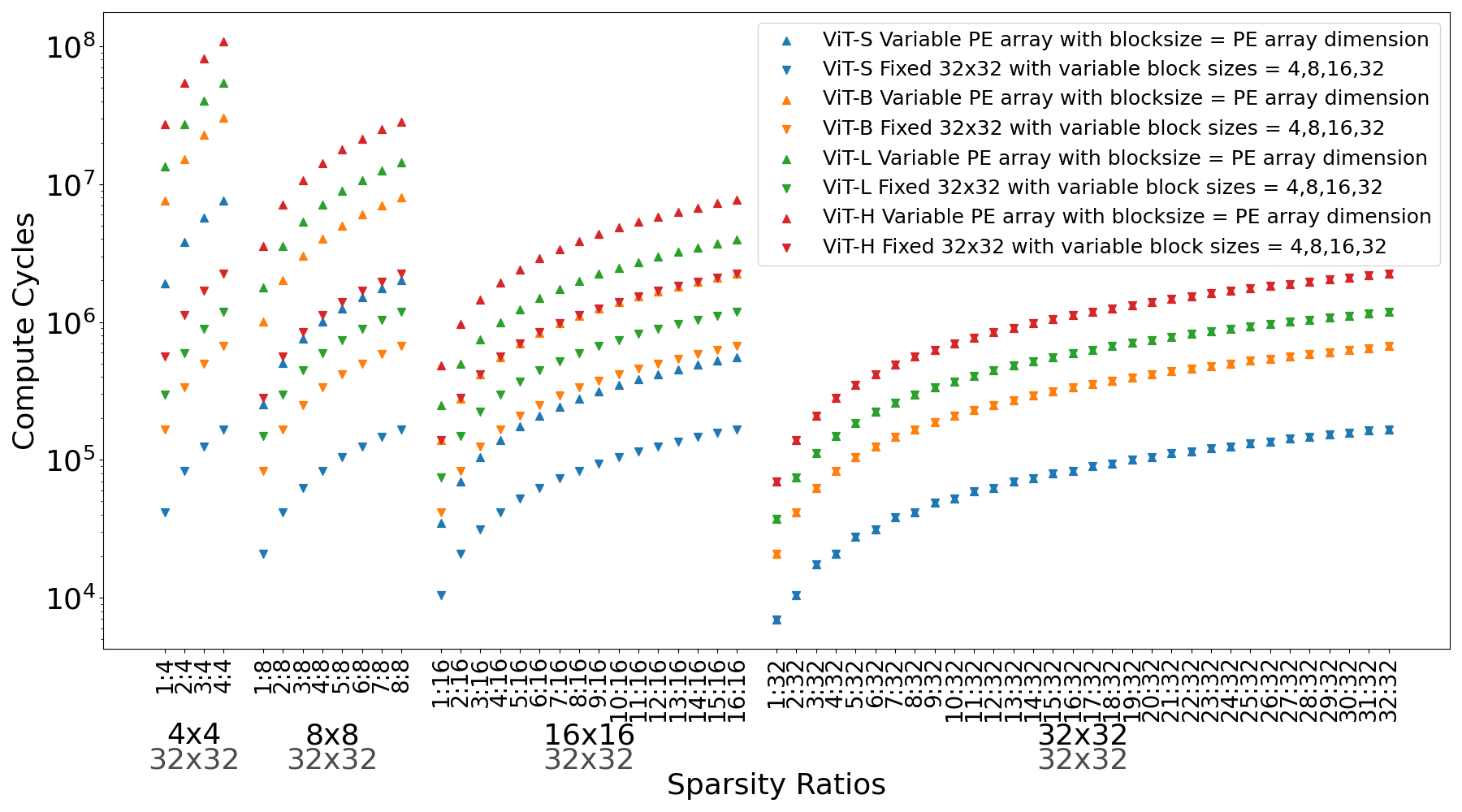}
  \caption{Compute cycle variation shown for Feed Forward layers of ViTs for different array sizes, sparsity ratios and block sizes.}
  \label{fig:sparsity-gemm-computecycles-overlap}
  \vspace{-0.17in}
\end{figure}

To study the effects of varying block sizes on performance, a study has been performed as shown in \autoref{fig:sparsity-gemm-computecycles-overlap}. The first set of runs has varying systolic array sizes (4x4, 8x8, 16x16, 32x32) and the block size considered is always equal to the systolic array dimension. Hence, sparsity ratio ranges only from $1:M$ to $M:M$. This is compared against the second set of runs that has a fixed array size (32x32) and the block size considered is varied ($M$ = 4, 8, 16 and 32). For each $M$, we shall have $M$ different values of $N$, thus resulting in $N$ different sparsity ratios. By increasing the block size, we get more finer granular control over the model, and utilizing the lower range/spectrum of $N:M$ values leads to better performances.


\section{Main Memory Integration}
\label{sec:ramulator}

One of the key limitations of \NAME v2 was the absence of a memory model with the systolic computation unit.  
The earlier version operated under the assumption that the data would somehow be readily available in the on-chip scratchpad.
This assumption overlooked the estimation of memory latency, stalls caused by DRAM bank conflicts and other pipeline bubbles in the datapath, critical in memory intensive workloads.

\NAME v3 addresses this limitation by integrating Ramulator~\cite{kim2015ramulator} memory model with the systolic array.
It enables memory datapath modeling by configuring the load/store queue size, the DRAM channels and the memory technology (i.e, DDR4, HBM, LPDDR4 etc.).
It also adds the memory latency, along with pipeline stalls to the ML inference latency.

We first introduce the primitives required for main memory integration in \autoref{subsec:rmltr_primitives}, 
followed by the steps to simulate the ML inference latency using Ramulator in \autoref{subsec:rmltr_flow}, 
and finally reporting some results including memory throughput and the overall execution latency in \autoref{subsec:rmltr_results}.

\subsection{Memory Integration Primitives}
\label{subsec:rmltr_primitives}
\subsubsection{Main Memory Model} 
A detailed memory model simulates the round-trip latency for each load and store memory transaction. 
Existing main memory models like Ramulator~\cite{kim2015ramulator,ramulatorv2}, DRAMSim3~\cite{dramsim3} etc. simulate different DRAM technologies while allowing the user to configure the memory controller. 
The model takes a memory demand request trace as an input.
Each trace entry comprises of the request cycle, memory address, and transaction type (read or write). 
The main memory model provides the round-trip latency for each transaction, along with overall runtime statistics including the memory throughput, row buffer hits/misses, bank conflicts etc. 
This statistics can be used to study the execution bottlenecks of each workload dataflows, memory controller configurations and main memory technologies.

\subsubsection{Finite Memory Request Queues}
\NAME v3 adds configurable memory request queues to hold pending memory transactions. 
Demand requests logged in the request queues are cleared upon transaction completion. 
Read requests are cleared once the data value is received, whereas write requests are cleared once it is logged in the memory controller. 
The finite size of these request queues stalls the accelerator when the pending queue is full. 
These queues help model memory stalls that impact execution runtime of memory-intensive workloads. 

\subsubsection{Memory delay modeling}
A systolic array operates on multiple data bits, which are fed from the input and the weight scratchpad SRAM buffers. 
\NAME v3 accounts for the memory load delay and triggers the systolic array only when each data is available in the scratchpad.
This adds runtime latency and pipeline bubbles during data streaming degrading the compute performance for certain dataflows. 
Exact modeling of data-plane components enables the user to evaluate dataflow, sparsity, and other workload optimizations.

\subsection{Memory Simulation Runtime}
\label{subsec:rmltr_flow}

The main memory simulation workflow is as follows: 

\textbf{Step 1: Accelerator memory request generation:}
The systolic array simulator is used to generate the memory demand transactions based on the sequence of data required to execute the workload. 
The \textit{ifmap, filter and ofmap} addresses are generated along with a cycle timestamp, which are sequenced to form a single memory trace.
Each entry contains a request cycle, a memory address and a transaction type. 

\textbf{Step 2: Evaluating the memory round-trip latency:}
We feed the generated memory trace to a memory simulator. We use Ramulator~\cite{kim2015ramulator} in our evaluation. 
The memory parameters like technology, number of ranks and channels, frequency etc. are set in a configuration file. 
The memory simulator reports the round-trip latency for each individual request.
The memory model also captures memory statistics like throughput, row-buffer hits and misses. 
The round-trip latency includes delays from memory bank conflicts and controller delays.

\textbf{Step 3: Simulating \NAME v3 with memory delays:}
The systolic array simulator is run again, but with a finite request queue and with the actual memory load/store latency. 
The data from the memory simulator is sent as input to \NAME v3. 
The execution stalls are calculated by enabling the finite memory request queues and the memory delay modeling described in \autoref{subsec:rmltr_primitives}.
Overall, the above workflow provides a realistic systolic array efficiency based on the memory block simulated for an ML accelerator. 

\begin{figure}[hbtp]
  \centering
  \includegraphics[width=\linewidth]{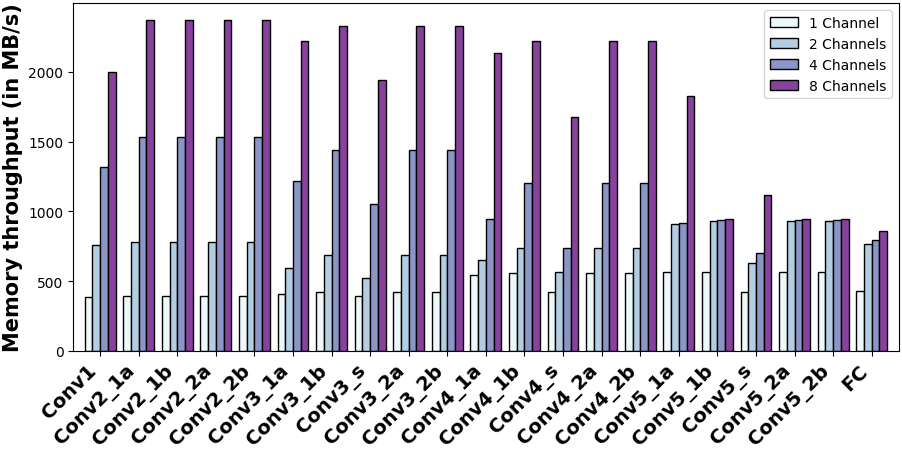}
  \caption{Impact of DRAM channels on memory throughput in Resnet18}
  \label{fig:bw_channel}
\end{figure}

\subsection{Memory Model Evaluation}
\label{subsec:rmltr_results}

\subsubsection{\NAME v3 and Memory Configuration}
\NAME v3 is run with the Google TPU configuration. The read and write request queues are individually configured to a size of $128$ entries.
The Ramulator simulates a DDR4 memory with 4Gb capacity for each channel with a $2400$ MHz speed.

\subsubsection{Impact of DRAM channels on memory throughput}
ML inference execution is known to benefit from multiple DRAM channels, especially for memory-intensive layers. 
We evaluate the impact of memory throughput for Resnet18 layers by tuning the number of DDR channels from 1 to 8 as shown in \autoref{fig:bw_channel}.
We observe a proportional increase in the memory throughput with the addition of more memory channels for initial layers, 
the performance of the last layers saturates at a channel size of 2. 
The larger ifmap and filter sizes of the initial layers benefit from a larger number of channels, with some layers reaching a bandwidth of more than 2000 MB/s. 
The $1\times1$ filters and smaller ifmaps reduce the memory throughput for later convolution and fully connected layers. 
While memory-intensive layers and streaming patterns benefit from multiple channels, compute-intensive or smaller layers fail to leverage multiple channels efficiently. 
Each memory channel also comes at an additional area cost for the memory controller and a power cost for parallel data loads.

\subsubsection{Stalls created by memory modeling }
\begin{figure}[hbtp]
  \centering
  \includegraphics[width=\linewidth]{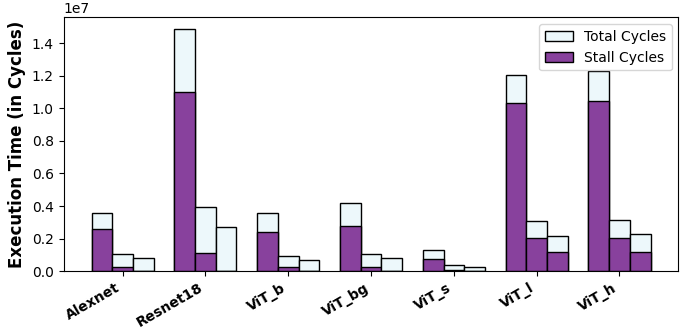}
  \caption{Impact of memory stalls on the overall ML inference latency. The three bars configure a read/write request queue of size 32, 128 and 512 respectively. }
  \label{fig:stall_cycle}
\end{figure}

Prior ML inference simulators did not model stalls emanating from the memory transactions. 
We define the number of \textit{stall cycles} as execution cycles when the systolic array is waiting on data in the scratchpad.  
\autoref{fig:stall_cycle} illustrates the number of stall cycles for different workloads as a fraction of the total execution cycles. 
The three bars represent the demand request queue size of $32$, $128$ and $512$ entries. 
The $32$ entry request queue certainly adds a lot of stall cycles as the systolic array easily fills a small queue with demand requests.
However, we see the percentage of stall cycles and the total execution cycles reduce with increasing entries. 
The average total cycles reduce by $3.76\times$ when the request queue size changes from 32 to 128 with a further improvement of $38\%$ with 512 entries.
Overall, the stall cycles due to memory simulation can delay the overall inference latency from 10 thousand cycles to a few million cycles depending on the workload and is important to consider while estimating the overall inference latency. The stall cycles in \autoref{fig:stall_cycle} show the percentage of unaccounted execution inefficiency in SCALE-Sim-v2 (which assumed a zero latency data transfer). DRAM modeling enhances the precision of total execution cycle estimates and optimizes overall systolic array utilization metrics.


\section{Data Layout in Memory}
\label{sec:layout}



In multi-core accelerators, multiple compute units share the common on-chip memory and process different workloads in parallel. When data is requested by multiple compute units located at different rows of the same bank within the on-chip SRAM, multiple cycles latency is required to bring required data out of the off-chip, termed as bank conflict slowdown. This slowdown can significantly degrade practical performance~\cite{tong2024featherreconfigurableacceleratordata}. To model this, we integrate precise layout modeling into \NAME v3.


In general, layout modeling includes two components:

\noindent \textbf{Layout Modeling:} Supports various data organizations in the multi-bank on-chip buffer.

\noindent \textbf{Memory Latency Evaluation with Layout Consideration:} Evaluates actual bank conflicts for accessing given data.

\subsection{Data Layout Modeling}

In modern multi-bank on-chip memory systems, global bandwidth is evenly distributed across memory banks. We model the multi-bank on-chip memory as a 2D array, where each row aggregates the row with the same index from all banks. This abstraction ensures the bandwidth of a single row matches the total on-chip bandwidth. Each bank has its own individual data accessing ports, enabling fine-grained data access and reducing bank conflicts.

\begin{figure}[t!]
  \vspace{-0.2in}
  \centering
  \includegraphics[width=\linewidth]{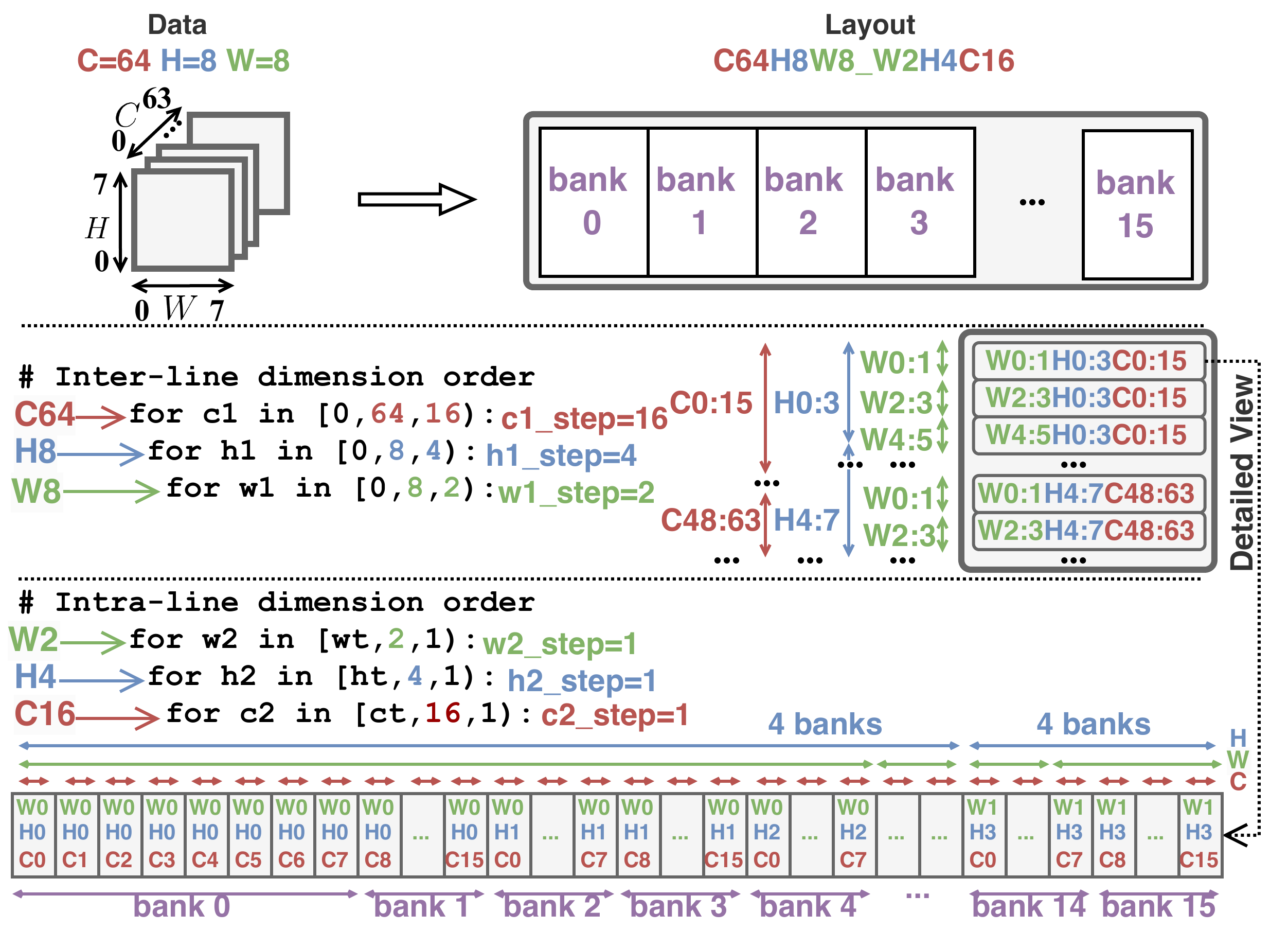}
  \caption{Overview of data layout modeling in multi-bank on-chip memory. We use nested loop to model data layout, where each line is an aggregation of the line with the same index from all banks. Each bank is 2 dimensional array offering limited number of ports for concurrent read/write access.}
  \label{fig:off_chip_mem_layout}
  \vspace{-0.17in}
\end{figure}

We define the data layout as the organization of data within the 2D array derived from the multi-bank memory. This includes determining the row and column positions of each data element. We term the order of row and column positions as inter-line dimension order and intra-line dimension order, respectively. Each dimension order is represented as nested loops, as shown in the \autoref{fig:off_chip_mem_layout}. 

\subsection{Memory Latency Evaluation}
\NAME v3 analyzes the latency for data requests from the compute array at the granularity of cycles. Taking a $C\times H \times W$ tensor from \autoref{fig:off_chip_mem_layout} as an example, when element $(c, h, w)$ is requested, the row-, column- and bank-index could be derived from the following equations.

\begin{align*}
    line_{id}= \lfloor c / c1\_step \rfloor \times \lfloor H / h1\_step \rfloor \times  \lfloor W / w1\_step \rfloor \\ +  \lfloor h / h1\_step \rfloor \times  \lfloor W / w1\_step \rfloor  \\ + \lfloor w / w1\_step \rfloor 
\end{align*}
\begin{align*}
    col_{id}= w \% w1\_step \times h1\_step \times  c1\_step \\ +  h \% h1\_step \times c2\_step  \\ +  c \% c1\_step 
\end{align*}
\begin{align*}
    bank_{id}= \lfloor \frac{col_{id}}{bandwidth\_per\_bank} \rfloor
\end{align*}

where $bandwidth\_per\_bank$ represents the maximal number of elements within a single line of a bank.

Given the $(line_{id}, col_{id}, bank_{id})$ for all data a certain dataflow might require, we could calculate the total number of lines in need of being accessed concurrently per cycle for each individual bank, yielding the following overall data accessing latency, assuming the total number of ports available for concurrent reading from each bank as $num\_ports$.

\begin{equation*}
    slowdown=\max_{i=0}^{num\_bank} \lceil  total\_rows\_bank_i / num\_ports\_bank_i \rceil
\end{equation*}

Among all banks in on-chip memory, the bank with the maximal bank conflicts determines the critical latency of providing requested data from on-chip memory.

\begin{figure}[t!]
  \vspace{-0.2in}
  \centering
  \includegraphics[width=\linewidth]{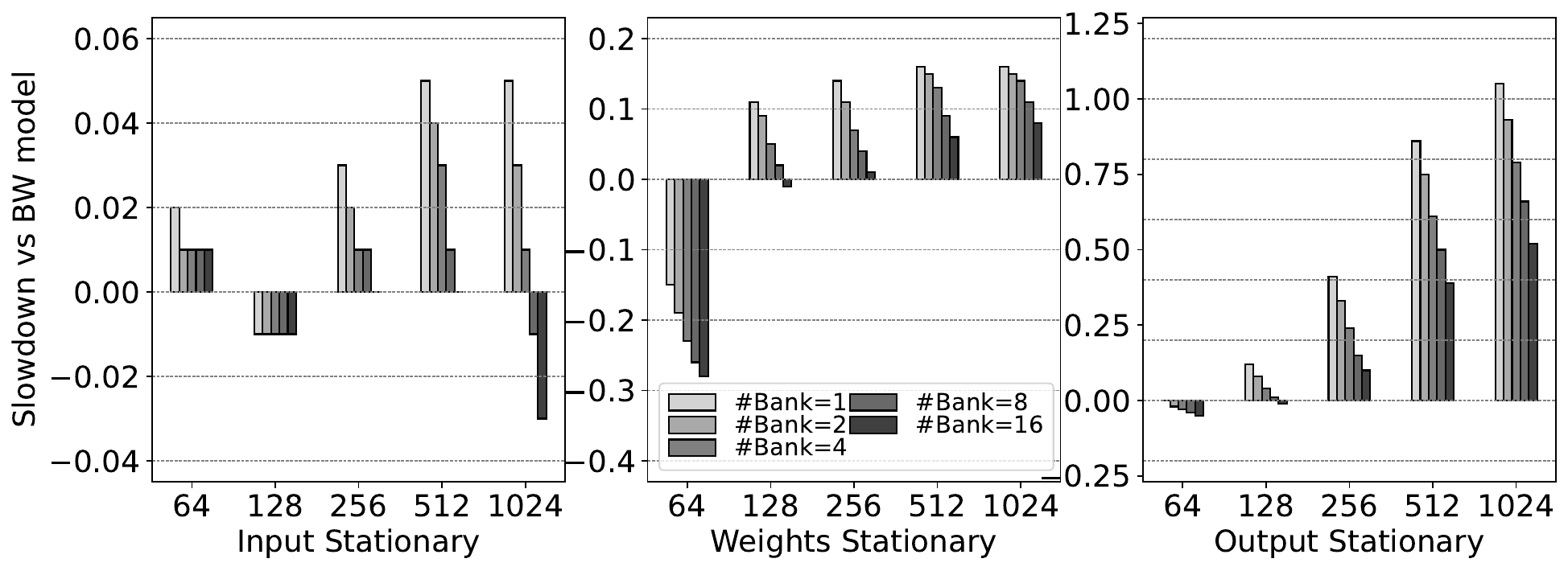}
  \caption{Slowdown of different dataflows under different (on-chip bandwidth, bank numbers). Area size: $128\times128$. Workload: ResNet18.}
  \label{fig:result_bandwidth_evaluation}
  \vspace{-0.1in}
\end{figure}
\begin{figure}[t!]
  \vspace{-0.1in}
  \centering
  \includegraphics[width=\linewidth]{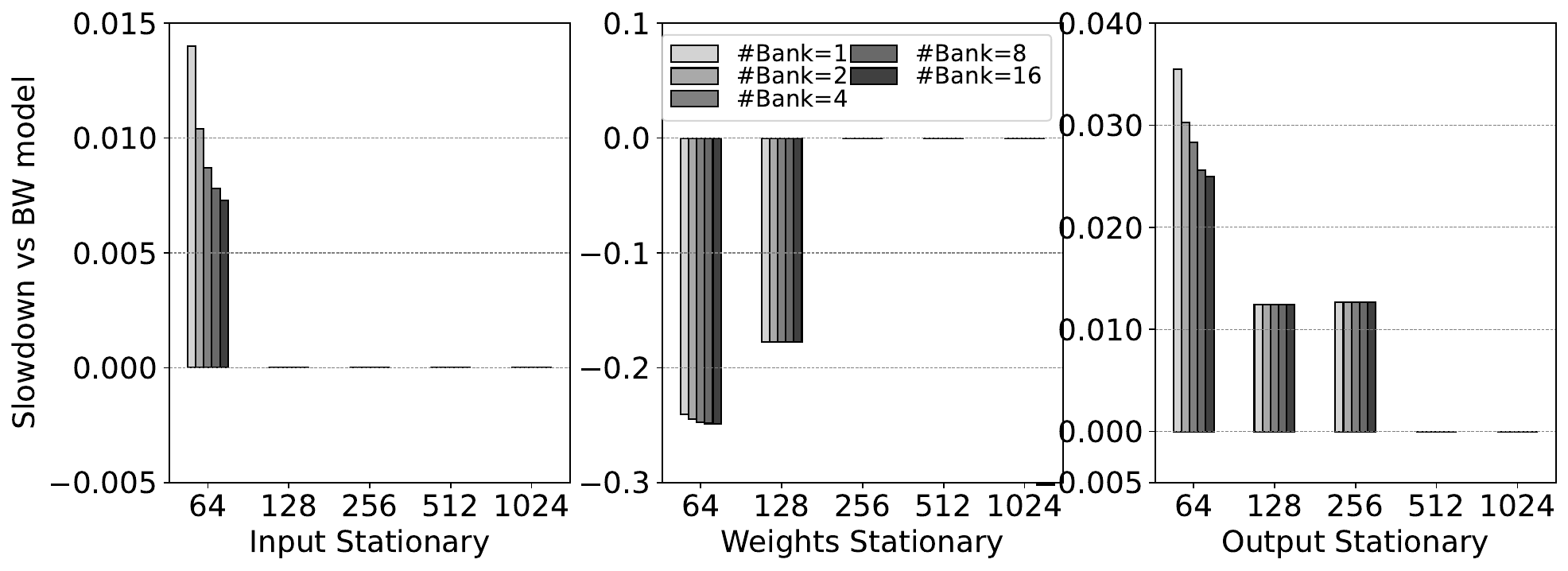}
  \caption{Slowdown of different dataflows under different (on-chip bandwidth, bank numbers). Area size: $128\times128$. Workload: ViT.}
  \label{fig:result_bandwidth_evaluation_vit}
  \vspace{-0.17in}
\end{figure}

The performance differences between realistic on-chip bank models and the original ideal bandwidth modeling in \NAME v2 are illustrated in \autoref{fig:result_bandwidth_evaluation} and \autoref{fig:result_bandwidth_evaluation_vit}. The figure normalizes latency under the realistic layout model against the pure bandwidth modeling in \NAME v2, highlighting the actual slowdown. Notably, given the same on-chip bandwidth, an increased number of banks consistently improves performance, as evidenced by the reduced slowdown with a higher bank count. This improvement is attributed to the finer-grained on-chip data access flexibility provided by additional banks, which effectively mitigates bank conflicts.

\section{Energy and Power Modeling}
\label{sec:accelergy}


This section presents the energy and power modeling of \NAME v3, achieved through the integration of Accelergy~\cite{wu2019accelergy}. It covers the framework overview (\autoref{subsec:accelergy_overview}) and details the integration methodology (\autoref{subsec:input}–~\autoref{subsec:PE_counts}).

\subsection{\NAME v3 Energy Modeling Framework Overview}
\label{subsec:accelergy_overview}
The \NAME v3 energy modeling framework comprises three key components: (1) unified user input, (2) trace files and action counts, and (3) performance and energy reports. Below is a detailed breakdown:

\textbf{Step 1: Architectural and Workload Input: }  
The neural network description remains unchanged across \NAME v3 and Accerlegy, serving as an independent input to generate trace files and cycle performance data. \NAME v3 automatically generates a YAML file for architectural input, combining user inputs with pre-defined systolic array component descriptions and energy estimation plugins to create the Energy Reference Table (ERT). Users can also customize component descriptions for greater flexibility.

\textbf{Step 2: Trace File and Action Counts: }
Accelergy distinguishes energy consumption for different action types (e.g., repeated vs. random memory accesses). To leverage this, \NAME v3 includes a feature to count action types in addition to generating trace files. The action counts file, along with performance metrics (latency, array utilization, and memory access counts), is produced by \NAME and used as input for Accelergy’s energy analysis.

\textbf{Step 3: Performance and Energy Output: }
After the Accelergy simulation completes, it generates a unified output that integrates energy estimations with \NAME v3’s cycle performance summary. This combined output provides insights into both timing predictability and energy efficiency, enabling users to optimize their accelerator design for a superior energy-delay product.

\begin{figure}[t!]
  \centering
  \includegraphics[width=\linewidth]{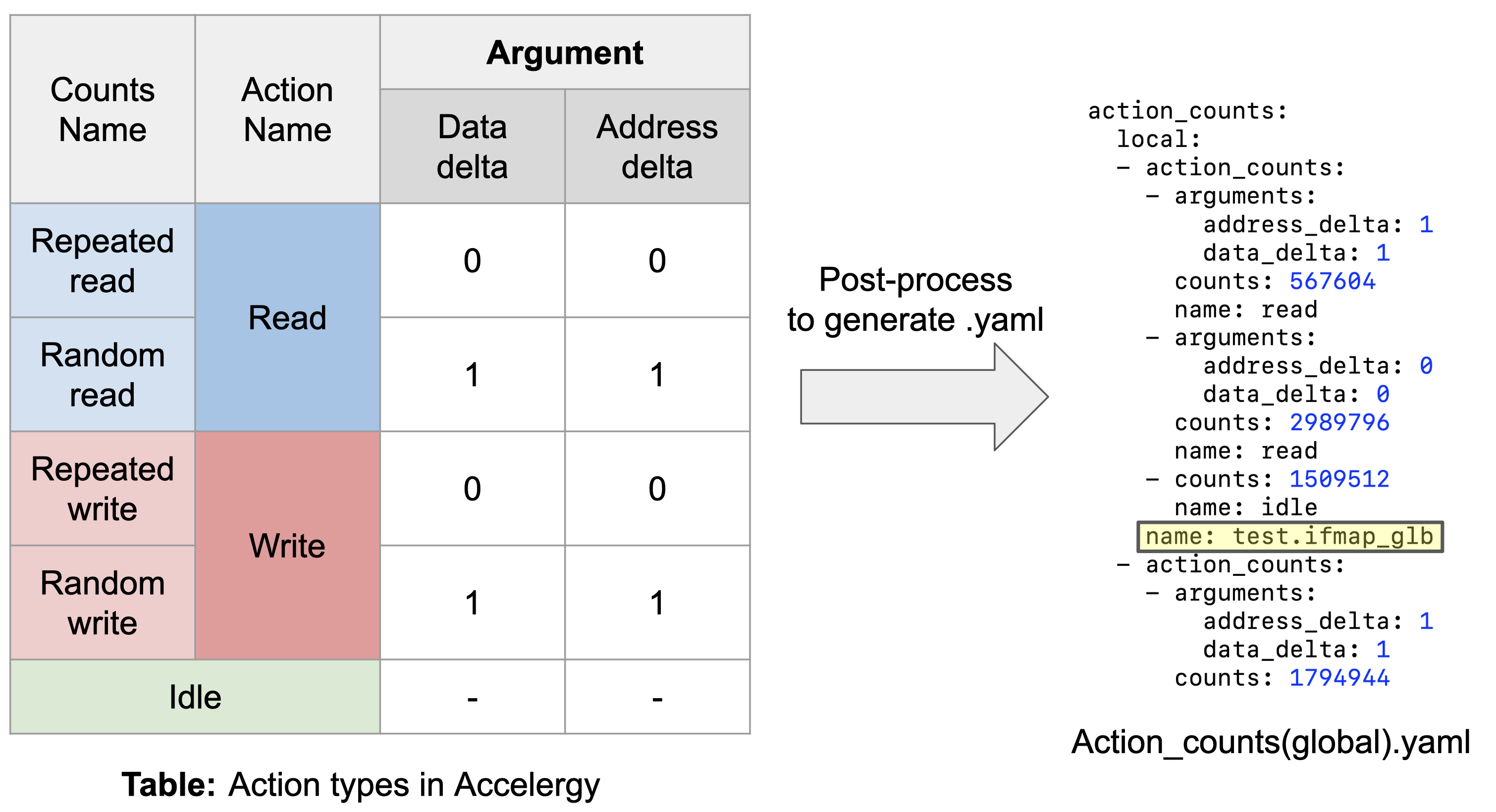}
  \caption{Translation table for action types in Accelergy with a snippet of the generated YAML file for memory action counts.}
  \label{fig:translation table}
  \vspace{-0.17in}
\end{figure}

\subsection{Input Integration of \NAME v3 and Accelergy}
\label{subsec:input}
The input files for \NAME v3 and Accelergy differ in abstraction levels. \NAME v3’s configuration input file specifies high-level configurations such as array shape, SRAM sizes, and dataflow, while Accelergy’s architecture YAML file describes lower-level details like hardware component types, memory bank counts, and data widths.
 
To ensure consistency between the two frameworks, we introduce a YAML file generator that extracts high-level systolic array configurations and extrapolates detailed architectural descriptions using a baseline YAML template, and outputs the final architecture.yaml file for energy modeling. The baseline template includes default components such as three register files and an integer MAC unit for each PE, along with three smart buffer SRAMs for input activations, weights, and partial sums. Users can customize component types by providing corresponding Accelergy-compatible descriptions.
 
 
\subsection{Repeated Access Lookup}
\label{susbec:repeated}

One of the key features enabling Accelergy's accurate energy estimates is its ability to distinguish between different action types, such as repeated and random memory accesses, which can differ in energy consumption by more than double. To support this, \NAME v3 generates an action count file as input for Accelergy. 

\NAME v3 introduces two tunable parameters: 'row size' and 'bank size.' The 'row size' parameter defines the block of data retrieved during each memory access, allowing for energy-efficient repeated reads when accessing consecutive addresses within the same block. The 'bank size' parameter accounts for multiple row buffers in a memory bank, enabling data reuse across cycles without reloading.



\subsection{Memory Action Counts}
\label{subsec:memory_count}
As discussed in ~\autoref{susbec:repeated}, the action-count summary file enables Accelergy to provide detailed architecture-level energy estimates. The dataflow between \NAME v3 and Accelergy occurs in two stages.

First, \NAME v3 generates memory transaction counts (idle, random-read/write, and repeated-read/write) for the ifmap, ofmap, and filter data using the implemented repeat count function. For example:
\begin{equation*}
    ifmap_{SRAM-idle}=cycles\times arraysize-counts
\end{equation*}
\begin{equation*}
    ifmap_{SRAM-random}=counts - repeat-counts
\end{equation*}
These counts are exported into a summary file.

In the second stage, this data is post-processed to create a YAML file for Accelergy’s energy estimation. Additional arguments, such as $data\Delta$ and $address\Delta$, indicate whether the corresponding wire switches are activated. Detailed parameter configurations are shown in \autoref{fig:translation table}. The resulting YAML file is used as the primary input for Accelergy after the \NAME v3 simulation completes.

\subsection{Processing Element Action Counts}
\label{subsec:PE_counts}

\textbf{MAC action counts.}
In \NAME v3, we define two basic MAC actions: \textit{MAC\_random} (normal operations with new data) and \textit{MAC\_constant} (no data change or computation). These are calculated as follows:  
\begin{equation*}
    MAC\_random = \#PEs \times cycles \times utilization
\end{equation*}
\begin{equation*}
    MAC\_constant = \#PEs \times cycles \times (1 - utilization)
\end{equation*}
Unused MAC units can leverage clock gating, modeled in Accelergy as \textit{MAC\_gated}, which consumes only static energy, reducing dynamic clock energy consumption. This optimization is incorporated in our energy evaluation.

\textbf{Scratchpad action counts.}
There are three scratchpad memory inside each PE, namely ifmap\_scratch\_pad, weights\_scratch\_pad, and psum\_scratch\_pad. Based on Accelergy, each scratchpad has two action types: \textit{write} and \textit{read}. We generate action counts for each scratchpad. For \textit{weight\_spad}, we model \textit{write} counts as the number of SRAM filter reads, \textit{read} counts as the number of MAC operations. For \textit{ifmap\_spad}, we model \textit{write} counts as the number of SRAM IFMAP reads, \textit{read} counts as the number of MAC operations. For \textit{psum\_spad}, we model both \textit{write} and \textit{read} as the number of MAC operations. The number of writes and reads of each scratchpad relates to dataflow. Input stationary will result in less number of \textit{imap\_spad write} action. Output stationary makes each MAC unit responsible for all the computations required for an OFMAP pixel. Weight stationary will result in less number of \textit{weight\_spad write} action. 

\begin{figure}[t!]
  \centering
  \includegraphics[width=\linewidth]{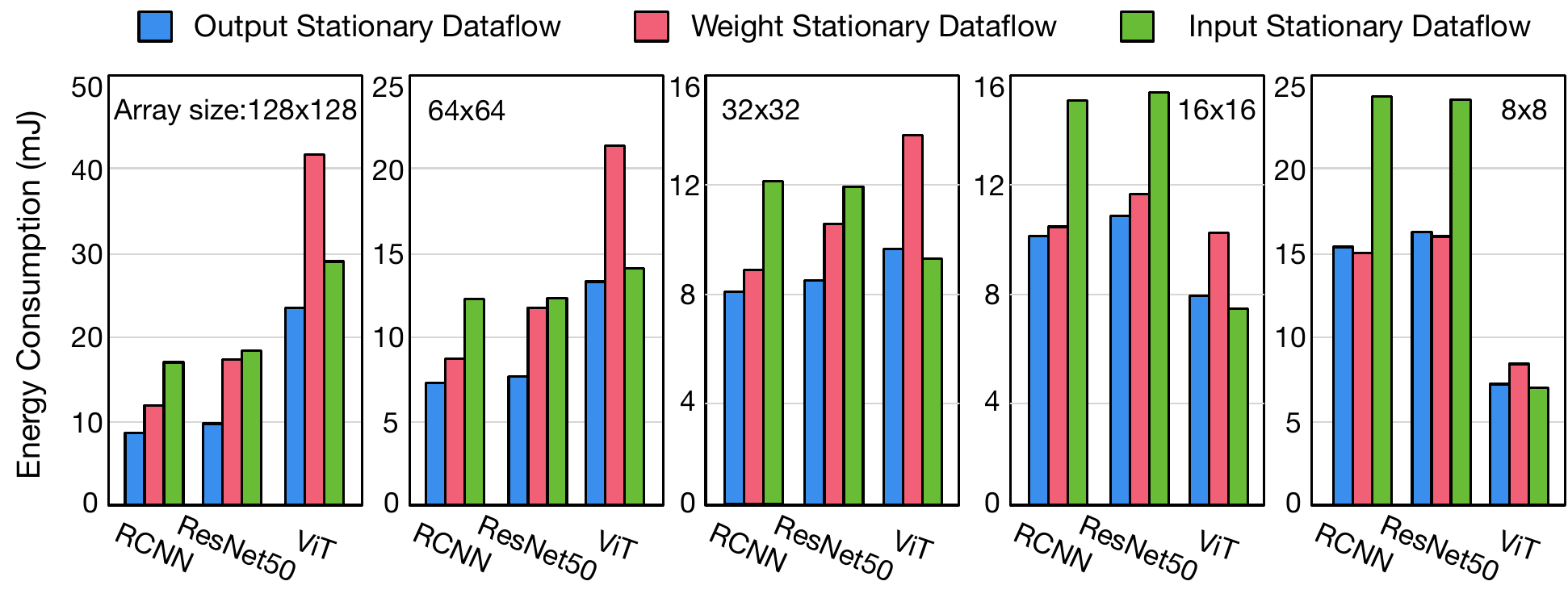}
  \caption{Energy consumption under different dataflow types and systolic array dimensions (128$\times$128, 64$\times$64, 32$\times$32, 16$\times$16, and 8$\times$8) across three workloads.}
  \label{fig:energy_result}
  \vspace{-0.17in}
\end{figure}

Putting all together, \autoref{fig:energy_result} presents an example of the energy consumption analysis for \NAME v3 when arrays of varying sizes execute the workload using different dataflow. OS outperforms the other two dataflows in almost every case. Compared with WS and IS, WS is preferable under smaller array sizes, while IS is preferable under larger array sizes. This analysis offers valuable insights into energy efficiency for designing diverse accelerators. Additionally, it can be seamlessly integrated with other features (e.g., Ramulator, sparsity, layout, etc.) to derive deeper design insights.
\section{Validation}

SCALE-Sim v3 builds on top of SCALE-Sim v2, Accelergy, and Ramulator, which are already validated against RTL of systolic array (cycle-accurate), Eyeriss chip (within 5\% error), and RTL of Micron DDR4 \cite{micronddr4} (cycle-accurate) respectively. We provide some additional validation information below:

\textbf{Sparsity.} We support two types of sparse accelerators:

\underline{Ampere fixed 2:4 sparsity.} SCALE-Sim v3 achieves 100\% accuracy compared to Sparse Tensor Core’s \cite{stc_report} report. While Sparseloop \cite{wu2022sparseloop} claims compute cycles are deterministic, memory stalls are not. SCALE-Sim v3 models cycle-accurate memory stalls, validated against RTL, making it more hardware-accurate.

\underline{Vegeta \cite{jeong2023vegeta} flexible N:M sparsity.} We validated 2:4 and 1:4 against Vegeta RTL with $ \leq 5\%$ error.

\textbf{Accelergy integration}: Using CIFAR-10 input data and 16-bit quantized weights from AlexNet and ResNet-18, we compare total energy consumption and relative energy breakdown (GLB, NoC, PE array) between PnR results (65nm) and SCALE-Sim v3, observing deviations of 4.6\% for Eyeriss and 4.8\% for a general systolic array (8x8 array, OS dataflow). Accelergy provides precise energy estimations for various system states (e.g., \textit{idle}, \textit{read\_rand}, \textit{read\_repeat}, \textit{write\_rand}, \textit{write\_repeat}, \textit{write\_cst\_data}) using unit energy values validated through RTL, synthesis, and PnR. SCALE-Sim v3 generates a cycle-accurate workload trace, which Accelergy processes to derive action counts for each state, enabling accurate energy and power calculations. This integration ensures a precise representation of system behavior, as shown in \autoref{tab:validation}:

\begin{table}[]
\centering
\resizebox{\columnwidth}{!}{%
\begin{tabular}{|
>{\columncolor[HTML]{FFFFFF}}l |
>{\columncolor[HTML]{FFFFFF}}l |
>{\columncolor[HTML]{FFFFFF}}l |
>{\columncolor[HTML]{FFFFFF}}l |}
\hline
\multicolumn{1}{|c|}{\cellcolor[HTML]{FFFFFF}{\color[HTML]{000000} \textbf{System State}}} & \multicolumn{1}{c|}{\cellcolor[HTML]{FFFFFF}{\color[HTML]{000000} \textbf{PnR Energy}}} & \multicolumn{1}{c|}{\cellcolor[HTML]{FFFFFF}{\color[HTML]{000000} \textbf{SCALE-Sim v3+Accelergy}}} & \multicolumn{1}{c|}{\cellcolor[HTML]{FFFFFF}{\color[HTML]{000000} \textbf{Error}}} \\ \hline
{\color[HTML]{000000} \textbf{Idle (clk gating)}}                                          & {\color[HTML]{000000} 12.3}                                                             & {\color[HTML]{000000} 12.6}                                                                                 & {\color[HTML]{000000} +2.4\%}                                                      \\ \hline
{\color[HTML]{000000} \textbf{Active}}                                                     & {\color[HTML]{000000} 315.8}                                                            & {\color[HTML]{000000} 308.5}                                                                                & {\color[HTML]{000000} -2.3\%}                                                      \\ \hline
{\color[HTML]{000000} \textbf{Power gating}}                                               & {\color[HTML]{000000} 4.7}                                                              & {\color[HTML]{000000} 4.9}                                                                                  & {\color[HTML]{000000} +4.3\%}                                                      \\ \hline
\end{tabular}%
}
\caption{Validation of Accelergy integration across system states (idle, active, power-gated)}
\label{tab:validation}
\end{table}

\textbf{Ramulator integration}: The read and write request queues are modeled to mimic the number of in-flight transactions in the AXI bus in the Micron DDR4 Verilog model \cite{micronddr4}.
\section{Comparison with \NAME v2}

\subsection{Simulation time comparison}

Simulation time overhead of \NAME v3 as compared to v2 is given in \autoref{tab:simulation}. The mean overhead of multi-core, 2:4 sparsity, 1:4 sparsity, Accelergy, Ramulator and Layout features are 2.29$\times$, 0.42$\times$, 0.29$\times$, 1.19$\times$, 2.13$\times$ and 16.03$\times$ respectively. Layout simulation overhead is large due to the detailed and accurate modeling of data layout in SRAM banks and bank conflicts.

\begin{table}[]
\centering
\resizebox{\columnwidth}{!}{%
\begin{tabular}{|
>{\columncolor[HTML]{FFFFFF}}l |
>{\columncolor[HTML]{FFFFFF}}l |
>{\columncolor[HTML]{FFFFFF}}l |
>{\columncolor[HTML]{FFFFFF}}l |
>{\columncolor[HTML]{FFFFFF}}l |
>{\columncolor[HTML]{FFFFFF}}l |
>{\columncolor[HTML]{FFFFFF}}l |}
\hline
\multicolumn{1}{|c|}{\cellcolor[HTML]{FFFFFF}{\color[HTML]{333333} \textbf{Workload}}} & \multicolumn{1}{c|}{\cellcolor[HTML]{FFFFFF}{\color[HTML]{333333} \textbf{\begin{tabular}[c]{@{}c@{}}Multi\\ core\end{tabular}}}} & \multicolumn{1}{c|}{\cellcolor[HTML]{FFFFFF}{\color[HTML]{333333} \textbf{\begin{tabular}[c]{@{}c@{}}Sparsity \\ 2:4\end{tabular}}}} & \multicolumn{1}{c|}{\cellcolor[HTML]{FFFFFF}{\color[HTML]{333333} \textbf{\begin{tabular}[c]{@{}c@{}}Sparsity \\ 1:4\end{tabular}}}} & \multicolumn{1}{c|}{\cellcolor[HTML]{FFFFFF}{\color[HTML]{333333} \textbf{\begin{tabular}[c]{@{}c@{}}Acce-\\ lergy\end{tabular}}}} & \multicolumn{1}{c|}{\cellcolor[HTML]{FFFFFF}{\color[HTML]{333333} \textbf{\begin{tabular}[c]{@{}c@{}}Ramu-\\ lator\end{tabular}}}} & \multicolumn{1}{c|}{\cellcolor[HTML]{FFFFFF}{\color[HTML]{333333} \textbf{Layout}}} \\ \hline
{\color[HTML]{333333} \textbf{Alexnet}}                                                & {\color[HTML]{333333} 1.83x}                                                                                                      & {\color[HTML]{333333} 0.29x}                                                                                                         & {\color[HTML]{333333} 0.14x}                                                                                                         & {\color[HTML]{333333} 1.44x}                                                                                                       & {\color[HTML]{333333} 1.19x}                                                                                                       & {\color[HTML]{333333} 22.62x}                                                       \\ \hline
{\color[HTML]{333333} \textbf{Resnet-18}}                                              & {\color[HTML]{333333} 2.06x}                                                                                                      & {\color[HTML]{333333} 0.30x}                                                                                                         & {\color[HTML]{333333} 0.30x}                                                                                                         & {\color[HTML]{333333} 1.09x}                                                                                                       & {\color[HTML]{333333} 1.74x}                                                                                                       & {\color[HTML]{333333} 3.6x}                                                         \\ \hline
{\color[HTML]{333333} \textbf{ViT-L}}                                                  & {\color[HTML]{333333} 2.59x}                                                                                                      & {\color[HTML]{333333} 0.47x}                                                                                                         & {\color[HTML]{333333} 0.32x}                                                                                                         & {\color[HTML]{333333} 1.06x}                                                                                                       & {\color[HTML]{333333} 2.03x}                                                                                                       & {\color[HTML]{333333} 18.88x}                                                       \\ \hline
{\color[HTML]{333333} \textbf{ViT-S}}                                                  & {\color[HTML]{333333} 2.67x}                                                                                                      & {\color[HTML]{333333} 0.62x}                                                                                                         & {\color[HTML]{333333} 0.39x}                                                                                                         & {\color[HTML]{333333} 1.17x}                                                                                                       & {\color[HTML]{333333} 3.57x}                                                                                                       & {\color[HTML]{333333} 19.01x}                                                       \\ \hline
\textbf{Mean}                                                                          & 2.29x                                                                                                                             & 0.42x                                                                                                                                & 0.29x                                                                                                                                & 1.19x                                                                                                                              & 2.13x                                                                                                                              & 16.03x                                                                              \\ \hline
\end{tabular}%
}
\caption{Simulation time overhead on a TPU-v2 like configuration}
\label{tab:simulation}
\end{table}

\subsection{Different designs found by \NAME v3 as compared to v2}

This section demonstrates the necessity for enhancements in SCALE-Sim v3 where it leads to a different design as compared to v2.

\begin{table*}[]
\centering
\resizebox{\textwidth}{!}{%
\begin{tabular}{|
>{\columncolor[HTML]{FFFFFF}}l |
>{\columncolor[HTML]{FFFFFF}}l 
>{\columncolor[HTML]{FFFFFF}}l 
>{\columncolor[HTML]{FFFFFF}}l |
>{\columncolor[HTML]{FFFFFF}}l 
>{\columncolor[HTML]{FFFFFF}}l 
>{\columncolor[HTML]{FFFFFF}}l |
>{\columncolor[HTML]{FFFFFF}}l 
>{\columncolor[HTML]{FFFFFF}}l 
>{\columncolor[HTML]{FFFFFF}}l |}
\hline
\multicolumn{1}{|c|}{\cellcolor[HTML]{FFFFFF}{\color[HTML]{333333} \textbf{Metric}}} & \multicolumn{3}{c|}{\cellcolor[HTML]{FFFFFF}{\color[HTML]{333333} \textbf{Resnet-50}}}                                                                                                         & \multicolumn{3}{c|}{\cellcolor[HTML]{FFFFFF}{\color[HTML]{333333} \textbf{RCNN}}}                                                                                                          & \multicolumn{3}{c|}{\cellcolor[HTML]{FFFFFF}{\color[HTML]{333333} \textbf{ViT-base}}}                                                                                                            \\ \hline
{\color[HTML]{333333} }                                                              & \multicolumn{1}{l|}{\cellcolor[HTML]{FFFFFF}{\color[HTML]{333333} \textbf{32x32}}}     & \multicolumn{1}{l|}{\cellcolor[HTML]{FFFFFF}{\color[HTML]{333333} \textbf{64x64}}}    & {\color[HTML]{333333} \textbf{128x128}}  & \multicolumn{1}{l|}{\cellcolor[HTML]{FFFFFF}{\color[HTML]{333333} \textbf{32x32}}}  & \multicolumn{1}{l|}{\cellcolor[HTML]{FFFFFF}{\color[HTML]{333333} \textbf{64x64}}}  & {\color[HTML]{333333} \textbf{128x128}}   & \multicolumn{1}{l|}{\cellcolor[HTML]{FFFFFF}{\color[HTML]{333333} \textbf{32x32}}}    & \multicolumn{1}{l|}{\cellcolor[HTML]{FFFFFF}{\color[HTML]{333333} \textbf{64x64}}}      & {\color[HTML]{333333} \textbf{128x128}}   \\ \hline
{\color[HTML]{333333} \textbf{Latency (cycles/layer)}}                               & \multicolumn{1}{l|}{\cellcolor[HTML]{FFFFFF}{\color[HTML]{333333} 98721}}     & \multicolumn{1}{l|}{\cellcolor[HTML]{FFFFFF}{\color[HTML]{333333} 35838}}    & {\color[HTML]{333333} 19501}    & \multicolumn{1}{l|}{\cellcolor[HTML]{FFFFFF}{\color[HTML]{333333} 126830}} & \multicolumn{1}{l|}{\cellcolor[HTML]{FFFFFF}{\color[HTML]{333333} 52243}}  & {\color[HTML]{333333} 29581}     & \multicolumn{1}{l|}{\cellcolor[HTML]{FFFFFF}{\color[HTML]{333333} 444970}}   & \multicolumn{1}{l|}{\cellcolor[HTML]{FFFFFF}{\color[HTML]{333333} 130601}}     & {\color[HTML]{333333} 68160}     \\ \hline
{\color[HTML]{333333} \textbf{Energy (mJ)}}                                          & \multicolumn{1}{l|}{\cellcolor[HTML]{FFFFFF}{\color[HTML]{333333} 6.91}}      & \multicolumn{1}{l|}{\cellcolor[HTML]{FFFFFF}{\color[HTML]{333333} 10.6}}     & {\color[HTML]{333333} 15.2}     & \multicolumn{1}{l|}{\cellcolor[HTML]{FFFFFF}{\color[HTML]{333333} 12.61}}  & \multicolumn{1}{l|}{\cellcolor[HTML]{FFFFFF}{\color[HTML]{333333} 28.36}}  & {\color[HTML]{333333} 29.25}     & \multicolumn{1}{l|}{\cellcolor[HTML]{FFFFFF}{\color[HTML]{333333} 11.02}}    & \multicolumn{1}{l|}{\cellcolor[HTML]{FFFFFF}{\color[HTML]{333333} 16.31}}      & {\color[HTML]{333333} 31.49}     \\ \hline
{\color[HTML]{333333} \textbf{EdP (cycles x mJ/layer)}}                              & \multicolumn{1}{l|}{\cellcolor[HTML]{FFFFFF}{\color[HTML]{333333} 682162.11}} & \multicolumn{1}{l|}{\cellcolor[HTML]{FFFFFF}{\color[HTML]{333333} 379882.8}} & {\color[HTML]{333333} 296415.2} & \multicolumn{1}{l|}{\cellcolor[HTML]{FFFFFF}{\color[HTML]{333333} 159932}} & \multicolumn{1}{l|}{\cellcolor[HTML]{FFFFFF}{\color[HTML]{333333} 148161}} & {\color[HTML]{333333} 865244.25} & \multicolumn{1}{l|}{\cellcolor[HTML]{FFFFFF}{\color[HTML]{333333} 490356.9}} & \multicolumn{1}{l|}{\cellcolor[HTML]{FFFFFF}{\color[HTML]{333333} 2130102.31}} & {\color[HTML]{333333} 2146358.4} \\ \hline
\end{tabular}%
}
\caption{Latency, Energy and EdP Comparison for 32x32, 64x64 and 128x128 arrays}
\label{tab:energy_comparison}
\end{table*}

\textbf{Energy. }SCALE-Sim v2 shows a 128×128 array is 6.53× faster than 32×32 for ViT-base, using only latency as a metric. However, v3 finds that 32×32 is 2.86× more energy-efficient due to lower leakage from better utilization. For EdP, 64×64 outperforms both 128×128 and 32×32 for ViT-base as shown in \autoref{tab:energy_comparison}.

\textbf{DRAM. } \NAME v2 shows a 21\% reduction in compute cycles for six ResNet18 layers using weight-stationary (WS) dataflow compared to output-stationary (OS). However, when factoring in DRAM stalls, OS dataflow exhibits 30.1\% lower execution cycles compared to WS, highlighting the critical role of detailed DRAM analysis. We see similar results in memory-intensive layers and for designs with smaller request queue sizes. We will add an analysis in the revision.

\textbf{Sparsity. } In \autoref{fig:sparsity-lws-resnet18-computecyclesplot}, a latency-constrained (250,000 cycles) design requires 3.00 MB on-chip memory for a dense core. Assuming 2:4 sparsity, SCALE-Sim v3 enables a sparse core with 768 kB, significantly reducing area.

\begin{table}[]
\centering
\resizebox{\columnwidth}{!}{%
\begin{tabular}{|
>{\columncolor[HTML]{FFFFFF}}l |
>{\columncolor[HTML]{FFFFFF}}l 
>{\columncolor[HTML]{FFFFFF}}l |
>{\columncolor[HTML]{FFFFFF}}l 
>{\columncolor[HTML]{FFFFFF}}l |}
\hline
\multicolumn{1}{|c|}{\cellcolor[HTML]{FFFFFF}{\color[HTML]{333333} \textbf{Dataflow}}} & \multicolumn{2}{c|}{\cellcolor[HTML]{FFFFFF}{\color[HTML]{333333} \textbf{Single core 128x128}}}                              & \multicolumn{2}{c|}{\cellcolor[HTML]{FFFFFF}{\color[HTML]{333333} \textbf{16 cores having 32x32 PEs}}}             \\ \hline
{\color[HTML]{333333} }                                                                & \multicolumn{1}{l|}{\cellcolor[HTML]{FFFFFF}{\color[HTML]{333333} \textbf{Latency}}} & {\color[HTML]{333333} \textbf{Energy}} & \multicolumn{1}{l|}{\cellcolor[HTML]{FFFFFF}{\color[HTML]{333333} \textbf{Latency}}} & {\color[HTML]{333333} \textbf{Energy}} \\ \hline
{\color[HTML]{333333} \textbf{Ratio ws/is}}                                            & \multicolumn{1}{l|}{\cellcolor[HTML]{FFFFFF}{\color[HTML]{333333} 1.87}}             & {\color[HTML]{333333} 0.71}            & \multicolumn{1}{l|}{\cellcolor[HTML]{FFFFFF}{\color[HTML]{333333} 1.14}}             & {\color[HTML]{333333} 0.70}            \\ \hline
\end{tabular}%
}
\caption{Latency and energy overhead of multi-core}
\label{tab:multi-core}
\end{table}

\textbf{Multi-core. } Comparing iso-compute designs (128×128 single-core vs. 16×32×32 multi-cores) for ViT-base, weight stationary outperforms input stationary by 1.87× (single-core) and 1.14× (multi-core) in latency. SCALE-Sim v2 favors weight stationary, but v3 allows further analysis before ruling out input stationary due to a small difference in latency, which performs 1.31× better in EdP for multi-core as shown in \autoref{tab:multi-core}.
\section{Related Work}

\textbf{Cycle accurate simulators.}
Cycle-accurate simulators are crucial tools for evaluating the performance and efficiency of AI accelerators, particularly those based on systolic arrays. Several existing simulators address various aspects of accelerator design as shown in \autoref{tab:related works}. uSystolic-SIM \cite{wu2022usystolic} is built on top of \NAME \cite{samajdar2018scale} which introduces a unique approach to cycle-level simulation by leveraging unary encoding to simplify computations, though it primarily targets byte-crawling architectures and does not extend to multi-core or sparse accelerators.

\textbf{Multi-core simulators.}
Multi-core systolic accelerators have gained traction for their ability to enhance performance in data-parallel workloads.  Tools like Timeloop \cite{parashar2019timeloop} offer analytical modeling for multi-core accelerators with basic energy estimation through Accelergy \cite{wu2019accelergy}, but they lack cycle-accurate simulation and data layout support. MuchiSim \cite{orenes2024muchisim} is another notable tool that simulates large-scale multi-core and multi-chip accelerators with detailed modeling of inter-core communication and memory hierarchies. However, it lacks compute modeling and relies on third-party tools.

\textbf{Sparse accelerator simulators.}
Sparse accelerators have gained prominence in recent times and various simulators have been developed to explore the design and performance of such accelerators. Sparseloop \cite{wu2022sparseloop} extends Timeloop’s capabilities by adding explicit support for sparse dataflows and mappings, enabling detailed exploration of sparsity patterns and their effects on accelerator performance. It models sparsity-related hardware optimizations including gating, skipping, and compression. However, it is analytical (modeling sparsity as a distribution) and lacks the support for cycle-accurate insights. STONNE \cite{munoz2021stonne} offers a cycle-accurate simulation of a specific unstructured sparse accelerator - SIGMA~\cite{qin2020sigma}.  However, it does not support multi-core architectures, data layout, or detailed memory modeling.

\section{Conclusion}
In this work, we introduced SCALE-Sim v3, a modular and cycle-accurate simulator designed to address the evolving needs of modern AI accelerator design and analysis. Our enhancements over SCALE-Sim v2 provide substantial improvements, including support for multi-core features, spatio-temporal partitioning, sparse matrix operations, memory interface modeling via Ramulator, energy/power estimation through Accelergy and on-chip data layout modeling. These features collectively enable comprehensive end-to-end system analysis, facilitating deeper insights into the performance, efficiency, and architectural trade-offs. We believe SCALE-Sim v3 will be a valuable tool for researchers and developers aiming to explore and optimize the design space of domain-specific AI accelerators. 



\section*{Acknowledgements}
This work was supported in part by ACE, one of seven centers in JUMP 2.0, a Semiconductor Research Corporation (SRC) program sponsored by DARPA.

\bibliographystyle{ieeetr}

\bibliography{main}

\begin{thebibliography}{10}

\bibitem{sevilla2022compute}
J.~Sevilla, L.~Heim, A.~Ho, T.~Besiroglu, M.~Hobbhahn, and P.~Villalobos, ``Compute trends across three eras of machine learning,'' in {\em 2022 International Joint Conference on Neural Networks (IJCNN)}, pp.~1--8, IEEE, 2022.

\bibitem{jouppi2023tpu}
N.~Jouppi, G.~Kurian, S.~Li, P.~Ma, R.~Nagarajan, L.~Nai, N.~Patil, S.~Subramanian, A.~Swing, B.~Towles, {\em et~al.}, ``Tpu v4: An optically reconfigurable supercomputer for machine learning with hardware support for embeddings,'' in {\em Proceedings of the 50th Annual International Symposium on Computer Architecture (ISCA)}, pp.~1--14, 2023.

\bibitem{Cerebras}
{Cerebras}, ``{Wafer Scale Engine (WSE-2) Dataset}.''
\newblock \url{https://f.hubspotusercontent30.net/hubfs/8968533/WSE-2%20Datasheet.pdf}.

\bibitem{firoozshahian2023mtia}
A.~Firoozshahian, J.~Coburn, R.~Levenstein, R.~Nattoji, A.~Kamath, O.~Wu, G.~Grewal, H.~Aepala, B.~Jakka, B.~Dreyer, {\em et~al.}, ``Mtia: First generation silicon targeting meta's recommendation systems,'' in {\em Proceedings of the 50th Annual International Symposium on Computer Architecture}, pp.~1--13, 2023.

\bibitem{lichtenau2022ai}
C.~Lichtenau, A.~Buyuktosunoglu, R.~Bertran, P.~Figuli, C.~Jacobi, N.~Papandreou, H.~Pozidis, A.~Saporito, A.~Sica, and E.~Tzortzatos, ``Ai accelerator on ibm telum processor: Industrial product,'' in {\em Proceedings of the 49th Annual International Symposium on Computer Architecture}, pp.~1012--1028, 2022.

\bibitem{reuther2020survey}
A.~Reuther, P.~Michaleas, M.~Jones, V.~Gadepally, S.~Samsi, and J.~Kepner, ``Survey of machine learning accelerators,'' in {\em 2020 IEEE high performance extreme computing conference (HPEC)}, pp.~1--12, IEEE, 2020.

\bibitem{feinberg2018enabling}
B.~Feinberg, U.~K.~R. Vengalam, N.~Whitehair, S.~Wang, and E.~Ipek, ``Enabling scientific computing on memristive accelerators,'' in {\em 2018 ACM/IEEE 45th Annual International Symposium on Computer Architecture (ISCA)}, pp.~367--382, IEEE, 2018.

\bibitem{sun2022burstz+}
G.~Sun, S.~Kang, and S.-W. Jun, ``Burstz+: Eliminating the communication bottleneck of scientific computing accelerators via accelerated compression,'' {\em ACM Transactions on Reconfigurable Technology and Systems (TRETS)}, vol.~15, no.~2, pp.~1--34, 2022.

\bibitem{bocco2019smurf}
A.~Bocco, Y.~Durand, and F.~De~Dinechin, ``Smurf: Scalar multiple-precision unum risc-v floating-point accelerator for scientific computing,'' in {\em Proceedings of the Conference for Next Generation Arithmetic 2019}, pp.~1--8, 2019.

\bibitem{weber2010comparing}
R.~Weber, A.~Gothandaraman, R.~J. Hinde, and G.~D. Peterson, ``Comparing hardware accelerators in scientific applications: A case study,'' {\em IEEE Transactions on Parallel and Distributed Systems}, vol.~22, no.~1, pp.~58--68, 2010.

\bibitem{ujaldon2016hpc}
M.~Ujald{\'o}n, ``Hpc accelerators with 3d memory,'' in {\em 2016 IEEE Intl Conference on Computational Science and Engineering (CSE) and IEEE Intl Conference on Embedded and Ubiquitous Computing (EUC) and 15th Intl Symposium on Distributed Computing and Applications for Business Engineering (DCABES)}, pp.~320--328, IEEE, 2016.

\bibitem{temam2012defect}
O.~Temam, ``A defect-tolerant accelerator for emerging high-performance applications,'' {\em ACM SIGARCH Computer Architecture News}, vol.~40, no.~3, pp.~356--367, 2012.

\bibitem{britt2017quantum}
K.~A. Britt, F.~A. Mohiyaddin, and T.~S. Humble, ``Quantum accelerators for high-performance computing systems,'' in {\em 2017 IEEE International Conference on Rebooting Computing (ICRC)}, pp.~1--7, IEEE, 2017.

\bibitem{li2022vector}
K.~Li, W.~Mao, J.~Zhou, B.~Li, Z.~Yang, S.~Yang, L.~Du, S.~Huang, and H.~Yu, ``A vector systolic accelerator for multi-precision floating-point high-performance computing,'' {\em IEEE Transactions on Circuits and Systems II: Express Briefs}, vol.~69, no.~10, pp.~4123--4127, 2022.

\bibitem{yang2022three}
L.~Yang, R.~M. Radway, Y.-H. Chen, T.~F. Wu, H.~Liu, E.~Ansari, V.~Chandra, S.~Mitra, and E.~Beign{\'e}, ``Three-dimensional stacked neural network accelerator architectures for ar/vr applications,'' {\em IEEE Micro}, vol.~42, no.~6, pp.~116--124, 2022.

\bibitem{sumbul2023fully}
H.~E. Sumbul, J.-s. Seo, D.~H. Morris, and E.~Beigne, ``A fully-digital and row-pipelined compute-in-memory neural network accelerator with soc-level benchmarking for ar/vr applications,'' {\em IEEE Micro}, 2023.

\bibitem{li2024fusion}
S.~Li, Y.~Zhao, C.~Li, B.~Guo, J.~Zhang, W.~Zhu, Z.~Ye, C.~Wan, and Y.~C. Lin, ``Fusion-3d: Integrated acceleration for instant 3d reconstruction and real-time rendering,'' in {\em 2024 57th IEEE/ACM International Symposium on Microarchitecture (MICRO)}, pp.~78--91, IEEE, 2024.

\bibitem{krishnan2022automatic}
S.~Krishnan, Z.~Wan, K.~Bhardwaj, P.~Whatmough, A.~Faust, S.~Neuman, G.-Y. Wei, D.~Brooks, and V.~J. Reddi, ``Automatic domain-specific soc design for autonomous unmanned aerial vehicles,'' in {\em 2022 55th IEEE/ACM International Symposium on Microarchitecture (MICRO)}, pp.~300--317, IEEE, 2022.

\bibitem{hao2024orianna}
Y.~Hao, Y.~Gan, B.~Yu, Q.~Liu, Y.~Han, Z.~Wan, and S.~Liu, ``Orianna: An accelerator generation framework for optimization-based robotic applications,'' in {\em Proceedings of the 29th ACM International Conference on Architectural Support for Programming Languages and Operating Systems, Volume 2}, pp.~813--829, 2024.

\bibitem{liu2022energy}
Q.~Liu, Z.~Wan, B.~Yu, W.~Liu, S.~Liu, and A.~Raychowdhury, ``An energy-efficient and runtime-reconfigurable fpga-based accelerator for robotic localization systems,'' in {\em 2022 IEEE Custom Integrated Circuits Conference (CICC)}, pp.~01--02, IEEE, 2022.

\bibitem{liu2021robotic}
S.~Liu, Z.~Wan, B.~Yu, and Y.~Wang, {\em Robotic computing on fpgas}.
\newblock Springer, 2021.

\bibitem{parashar2019timeloop}
A.~Parashar, P.~Raina, Y.~S. Shao, Y.-H. Chen, V.~A. Ying, A.~Mukkara, R.~Venkatesan, B.~Khailany, S.~W. Keckler, and J.~Emer, ``Timeloop: A systematic approach to dnn accelerator evaluation,'' in {\em 2019 IEEE international symposium on performance analysis of systems and software (ISPASS)}, pp.~304--315, IEEE, 2019.

\bibitem{munoz2021stonne}
F.~Mu{\~n}oz-Mart{\'\i}nez, J.~L. Abell{\'a}n, M.~E. Acacio, and T.~Krishna, ``Stonne: Enabling cycle-level microarchitectural simulation for dnn inference accelerators,'' in {\em 2021 IEEE International Symposium on Workload Characterization (IISWC)}, pp.~201--213, IEEE, 2021.

\bibitem{samajdar2018scale}
A.~Samajdar, Y.~Zhu, P.~Whatmough, M.~Mattina, and T.~Krishna, ``Scale-sim: Systolic cnn accelerator simulator,'' {\em arXiv preprint arXiv:1811.02883}, 2018.

\bibitem{kwon2020maestro}
H.~Kwon, P.~Chatarasi, V.~Sarkar, T.~Krishna, M.~Pellauer, and A.~Parashar, ``Maestro: A data-centric approach to understand reuse, performance, and hardware cost of dnn mappings,'' {\em IEEE micro}, vol.~40, no.~3, pp.~20--29, 2020.

\bibitem{wu2022usystolic}
D.~Wu and J.~San~Miguel, ``usystolic: Byte-crawling unary systolic array,'' in {\em 2022 IEEE International Symposium on High-Performance Computer Architecture (HPCA)}, pp.~12--24, IEEE, 2022.

\bibitem{chen2016eyeriss}
Y.-H. Chen, T.~Krishna, J.~S. Emer, and V.~Sze, ``Eyeriss: An energy-efficient reconfigurable accelerator for deep convolutional neural networks,'' {\em IEEE journal of solid-state circuits}, vol.~52, no.~1, pp.~127--138, 2016.

\bibitem{kwon2019understanding}
H.~Kwon, P.~Chatarasi, M.~Pellauer, A.~Parashar, V.~Sarkar, and T.~Krishna, ``Understanding reuse, performance, and hardware cost of dnn dataflow: A data-centric approach,'' in {\em Proceedings of the 52nd Annual IEEE/ACM International Symposium on Microarchitecture}, pp.~754--768, 2019.

\bibitem{muralimanohar2009cacti}
N.~Muralimanohar, R.~Balasubramonian, and N.~P. Jouppi, ``Cacti 6.0: A tool to model large caches,'' {\em HP laboratories}, vol.~27, p.~28, 2009.

\bibitem{qin2020sigma}
E.~Qin, A.~Samajdar, H.~Kwon, V.~Nadella, S.~Srinivasan, D.~Das, B.~Kaul, and T.~Krishna, ``Sigma: A sparse and irregular gemm accelerator with flexible interconnects for dnn training,'' in {\em 2020 IEEE International Symposium on High Performance Computer Architecture (HPCA)}, pp.~58--70, IEEE, 2020.

\bibitem{wu2019accelergy}
Y.~N. Wu, J.~S. Emer, and V.~Sze, ``Accelergy: An architecture-level energy estimation methodology for accelerator designs,'' in {\em 2019 IEEE/ACM International Conference on Computer-Aided Design (ICCAD)}, pp.~1--8, IEEE, 2019.

\bibitem{balasubramonian2017cacti}
R.~Balasubramonian, A.~B. Kahng, N.~Muralimanohar, A.~Shafiee, and V.~Srinivas, ``Cacti 7: New tools for interconnect exploration in innovative off-chip memories,'' {\em ACM Transactions on Architecture and Code Optimization (TACO)}, vol.~14, no.~2, pp.~1--25, 2017.

\bibitem{kim2015ramulator}
Y.~Kim, W.~Yang, and O.~Mutlu, ``Ramulator: A fast and extensible dram simulator,'' {\em IEEE Computer architecture letters}, vol.~15, no.~1, pp.~45--49, 2015.

\bibitem{tong2024featherreconfigurableacceleratordata}
J.~Tong, A.~Itagi, P.~Chatarasi, and T.~Krishna, ``Feather: A reconfigurable accelerator with data reordering support for low-cost on-chip dataflow switching,'' 2024.

\bibitem{kim2022ark}
J.~Kim, G.~Lee, S.~Kim, G.~Sohn, M.~Rhu, J.~Kim, and J.~H. Ahn, ``Ark: Fully homomorphic encryption accelerator with runtime data generation and inter-operation key reuse,'' in {\em 2022 55th IEEE/ACM International Symposium on Microarchitecture (MICRO)}, pp.~1237--1254, IEEE, 2022.

\bibitem{ma2023camj}
T.~Ma, Y.~Feng, X.~Zhang, and Y.~Zhu, ``Camj: Enabling system-level energy modeling and architectural exploration for in-sensor visual computing,'' in {\em Proceedings of the 50th Annual International Symposium on Computer Architecture}, pp.~1--14, 2023.

\bibitem{guo2023cambricon}
H.~Guo, Y.~Zhao, Z.~Li, Y.~Hao, C.~Liu, X.~Song, X.~Li, Z.~Du, R.~Zhang, Q.~Guo, {\em et~al.}, ``Cambricon-u: A systolic random increment memory architecture for unary computing,'' in {\em Proceedings of the 56th Annual IEEE/ACM International Symposium on Microarchitecture}, pp.~424--437, 2023.

\bibitem{samajdar2022self}
A.~Samajdar, E.~Qin, M.~Pellauer, and T.~Krishna, ``Self adaptive reconfigurable arrays (sara) learning flexible gemm accelerator configuration and mapping-space using ml,'' in {\em Proceedings of the 59th ACM/IEEE Design Automation Conference}, pp.~583--588, 2022.

\bibitem{liu2023deja}
Z.~Liu, J.~Wang, T.~Dao, T.~Zhou, B.~Yuan, Z.~Song, A.~Shrivastava, C.~Zhang, Y.~Tian, C.~Re, {\em et~al.}, ``Deja vu: Contextual sparsity for efficient llms at inference time,'' in {\em International Conference on Machine Learning}, pp.~22137--22176, PMLR, 2023.

\bibitem{gale2019state}
T.~Gale, E.~Elsen, and S.~Hooker, ``The state of sparsity in deep neural networks,'' {\em arXiv preprint arXiv:1902.09574}, 2019.

\bibitem{liu2015sparse}
B.~Liu, M.~Wang, H.~Foroosh, M.~Tappen, and M.~Pensky, ``Sparse convolutional neural networks,'' in {\em Proceedings of the IEEE conference on computer vision and pattern recognition}, pp.~806--814, 2015.

\bibitem{gale2023megablocks}
T.~Gale, D.~Narayanan, C.~Young, and M.~Zaharia, ``Megablocks: Efficient sparse training with mixture-of-experts,'' {\em Proceedings of Machine Learning and Systems}, vol.~5, pp.~288--304, 2023.

\bibitem{shen2023efficient}
H.~Shen, H.~Meng, B.~Dong, Z.~Wang, O.~Zafrir, Y.~Ding, Y.~Luo, H.~Chang, Q.~Gao, Z.~Wang, {\em et~al.}, ``An efficient sparse inference software accelerator for transformer-based language models on cpus,'' {\em arXiv preprint arXiv:2306.16601}, 2023.

\bibitem{giannoula2022towards}
C.~Giannoula, I.~Fernandez, J.~G{\'o}mez-Luna, N.~Koziris, G.~Goumas, and O.~Mutlu, ``Towards efficient sparse matrix vector multiplication on real processing-in-memory architectures,'' {\em ACM SIGMETRICS Performance Evaluation Review}, vol.~50, no.~1, pp.~33--34, 2022.

\bibitem{ampere}
``{NVIDIA Ampere Architecture}.'' \url{https://www.nvidia.com/en-in/data-center/ampere-architecture/}, 2020.

\bibitem{jeong2023vegeta}
G.~Jeong, S.~Damani, A.~R. Bambhaniya, E.~Qin, C.~J. Hughes, S.~Subramoney, H.~Kim, and T.~Krishna, ``Vegeta: Vertically-integrated extensions for sparse/dense gemm tile acceleration on cpus,'' in {\em 2023 IEEE International Symposium on High-Performance Computer Architecture (HPCA)}, pp.~259--272, IEEE, 2023.

\bibitem{kanellopoulos2019smash}
K.~Kanellopoulos, N.~Vijaykumar, C.~Giannoula, R.~Azizi, S.~Koppula, N.~M. Ghiasi, T.~Shahroodi, J.~G. Luna, and O.~Mutlu, ``Smash: Co-designing software compression and hardware-accelerated indexing for efficient sparse matrix operations,'' in {\em Proceedings of the 52nd annual IEEE/ACM international symposium on microarchitecture}, pp.~600--614, 2019.

\bibitem{zhou2018cambricon}
X.~Zhou, Z.~Du, Q.~Guo, S.~Liu, C.~Liu, C.~Wang, X.~Zhou, L.~Li, T.~Chen, and Y.~Chen, ``Cambricon-s: Addressing irregularity in sparse neural networks through a cooperative software/hardware approach,'' in {\em 2018 51st Annual IEEE/ACM International Symposium on Microarchitecture (MICRO)}, pp.~15--28, IEEE, 2018.

\bibitem{tpuv5}
{Google}, ``{TPUv5e}.''
\newblock \url{https://cloud.google.com/tpu/docs/v5e}.

\bibitem{jun2017hbm}
H.~Jun, J.~Cho, K.~Lee, H.-Y. Son, K.~Kim, H.~Jin, and K.~Kim, ``Hbm (high bandwidth memory) dram technology and architecture,'' in {\em 2017 IEEE International Memory Workshop (IMW)}, pp.~1--4, IEEE, 2017.

\bibitem{mutlu2017rowhammer}
O.~Mutlu, ``The rowhammer problem and other issues we may face as memory becomes denser,'' in {\em Design, Automation \& Test in Europe Conference \& Exhibition (DATE), 2017}, pp.~1116--1121, IEEE, 2017.

\bibitem{mutlu2019rowhammer}
O.~Mutlu and J.~S. Kim, ``Rowhammer: A retrospective,'' {\em IEEE Transactions on Computer-Aided Design of Integrated Circuits and Systems}, vol.~39, no.~8, pp.~1555--1571, 2019.

\bibitem{kung1979systolic}
H.~T. Kung and C.~E. Leiserson, ``Systolic arrays (for vlsi),'' in {\em Sparse Matrix Proceedings 1978}, vol.~1, pp.~256--282, Society for industrial and applied mathematics Philadelphia, PA, USA, 1979.

\bibitem{kung1982systolic}
H.-T. Kung, {\em Why systolic architecture?}
\newblock Design Research Center, Carnegie-Mellon University, 1982.

\bibitem{mead1980introduction}
C.~Mead and L.~Conway, ``Introduction to vlsi systems,'' 1980.

\bibitem{saptalakar2013design}
B.~K. Saptalakar, D.~Kale, M.~Rachannavar, and M.~Pavankumar, ``Design and implementation of vlsi systolic array multiplier for dsp applications,'' {\em International Journal of Scientific Engineering and Technology}, vol.~2, no.~3, pp.~156--159, 2013.

\bibitem{chiper2002systolic}
D.-F. Chiper, M.~Swamy, M.~O. Ahmad, and T.~Stouraitis, ``A systolic array architecture for the discrete sine transform,'' {\em IEEE transactions on signal processing}, vol.~50, no.~9, pp.~2347--2354, 2002.

\bibitem{quinton1983systematic}
P.~Quinton, {\em The systematic design of systolic arrays}.
\newblock PhD thesis, INRIA, 1983.

\bibitem{johnson1993general}
K.~T. Johnson, A.~R. Hurson, and B.~Shirazi, ``General-purpose systolic arrays,'' {\em Computer}, vol.~26, no.~11, pp.~20--31, 1993.

\bibitem{openai_compute}
{OpenAI}, ``{AI and compute}.''
\newblock \url{https://openai.com/index/ai-and-compute/}.

\bibitem{dubey2024llama}
A.~Dubey, A.~Jauhri, A.~Pandey, A.~Kadian, A.~Al-Dahle, A.~Letman, A.~Mathur, A.~Schelten, A.~Yang, A.~Fan, {\em et~al.}, ``The llama 3 herd of models,'' {\em arXiv preprint arXiv:2407.21783}, 2024.

\bibitem{team2023gemini}
G.~Team, R.~Anil, S.~Borgeaud, J.-B. Alayrac, J.~Yu, R.~Soricut, J.~Schalkwyk, A.~M. Dai, A.~Hauth, K.~Millican, {\em et~al.}, ``Gemini: a family of highly capable multimodal models,'' {\em arXiv preprint arXiv:2312.11805}, 2023.

\bibitem{achiam2023gpt}
J.~Achiam, S.~Adler, S.~Agarwal, L.~Ahmad, I.~Akkaya, F.~L. Aleman, D.~Almeida, J.~Altenschmidt, S.~Altman, S.~Anadkat, {\em et~al.}, ``Gpt-4 technical report,'' {\em arXiv preprint arXiv:2303.08774}, 2023.

\bibitem{zheng2025genad}
W.~Zheng, R.~Song, X.~Guo, C.~Zhang, and L.~Chen, ``Genad: Generative end-to-end autonomous driving,'' in {\em European Conference on Computer Vision}, pp.~87--104, Springer, 2025.

\bibitem{hu2023planning}
Y.~Hu, J.~Yang, L.~Chen, K.~Li, C.~Sima, X.~Zhu, S.~Chai, S.~Du, T.~Lin, W.~Wang, {\em et~al.}, ``Planning-oriented autonomous driving,'' in {\em Proceedings of the IEEE/CVF Conference on Computer Vision and Pattern Recognition}, pp.~17853--17862, 2023.

\bibitem{atakishiyev2024explainable}
S.~Atakishiyev, M.~Salameh, H.~Yao, and R.~Goebel, ``Explainable artificial intelligence for autonomous driving: A comprehensive overview and field guide for future research directions,'' {\em IEEE Access}, 2024.

\bibitem{rahman2023ambiguous}
A.~Rahman, J.~M.~J. Valanarasu, I.~Hacihaliloglu, and V.~M. Patel, ``Ambiguous medical image segmentation using diffusion models,'' in {\em Proceedings of the IEEE/CVF conference on computer vision and pattern recognition}, pp.~11536--11546, 2023.

\bibitem{zhang2022contrastive}
Y.~Zhang, H.~Jiang, Y.~Miura, C.~D. Manning, and C.~P. Langlotz, ``Contrastive learning of medical visual representations from paired images and text,'' in {\em Machine Learning for Healthcare Conference}, pp.~2--25, PMLR, 2022.

\bibitem{gao2018human}
Y.~Gao, X.~Xiang, N.~Xiong, B.~Huang, H.~J. Lee, R.~Alrifai, X.~Jiang, and Z.~Fang, ``Human action monitoring for healthcare based on deep learning,'' {\em Ieee Access}, vol.~6, pp.~52277--52285, 2018.

\bibitem{nvidia-dgx}
{Nvidia}, ``{DGX GH200 for Large Memory AI Supercomputer}.''
\newblock \url{https://www.nvidia.com/en-in/data-center/dgx-gh200/}.

\bibitem{talpes2022dojo}
E.~Talpes, D.~Williams, and D.~D. Sarma, ``Dojo: The microarchitecture of tesla’s exa-scale computer,'' in {\em 2022 IEEE Hot Chips 34 Symposium (HCS)}, pp.~1--28, IEEE Computer Society, 2022.

\bibitem{seo2024versa}
J.~Seo and J.~Kong, ``Versa: Versatile systolic array architecture for sparse and dense matrix multiplications,'' {\em Electronics}, vol.~13, no.~8, p.~1500, 2024.

\bibitem{ramulatorv2}
H.~Luo, Y.~C. Tu{\u{g}}rul, F.~N. Bostanc{\i}, A.~Olgun, A.~G. Ya{\u{g}}l{\i}k{\c{c}}{\i}, and O.~Mutlu, ``Ramulator 2.0: A modern, modular, and extensible dram simulator,'' {\em IEEE Computer Architecture Letters}, vol.~23, no.~1, pp.~112--116, 2023.

\bibitem{dramsim3}
S.~Li, Z.~Yang, D.~Reddy, A.~Srivastava, and B.~Jacob, ``Dramsim3: A cycle-accurate, thermal-capable dram simulator,'' {\em IEEE Computer Architecture Letters}, vol.~19, no.~2, pp.~106--109, 2020.

\bibitem{shao2019simba}
Y.~S. Shao, J.~Clemons, R.~Venkatesan, B.~Zimmer, M.~Fojtik, N.~Jiang, B.~Keller, A.~Klinefelter, N.~Pinckney, P.~Raina, {\em et~al.}, ``Simba: Scaling deep-learning inference with multi-chip-module-based architecture,'' in {\em Proceedings of the 52nd Annual IEEE/ACM International Symposium on Microarchitecture}, pp.~14--27, 2019.

\bibitem{symons2022towards}
A.~Symons, L.~Mei, S.~Colleman, P.~Houshmand, S.~Karl, and M.~Verhelst, ``Towards heterogeneous multi-core accelerators exploiting fine-grained scheduling of layer-fused deep neural networks,'' {\em arXiv preprint arXiv:2212.10612}, 2022.

\bibitem{nandakumar2022accelerating}
A.~Nandakumar, S.~Shao, and B.~Nikolic, ``Accelerating deep learning on heterogenous architectures,'' 2022.

\bibitem{spantidi2022targeting}
O.~Spantidi, G.~Zervakis, S.~Alsalamin, I.~Roman-Ballesteros, J.~Henkel, H.~Amrouch, and I.~Anagnostopoulos, ``Targeting dnn inference via efficient utilization of heterogeneous precision dnn accelerators,'' {\em IEEE Transactions on Emerging Topics in Computing}, vol.~11, no.~1, pp.~112--125, 2022.

\bibitem{lau1994chip}
J.~H. Lau, {\em Chip on board: technology for multichip modules}.
\newblock Springer Science \& Business Media, 1994.

\bibitem{doane2013multichip}
D.~A. Doane and P.~Franzon, {\em Multichip module technologies and alternatives: the basics}.
\newblock Springer Science \& Business Media, 2013.

\bibitem{arunkumar2017mcm}
A.~Arunkumar, E.~Bolotin, B.~Cho, U.~Milic, E.~Ebrahimi, O.~Villa, A.~Jaleel, C.-J. Wu, and D.~Nellans, ``Mcm-gpu: Multi-chip-module gpus for continued performance scalability,'' {\em ACM SIGARCH Computer Architecture News}, vol.~45, no.~2, pp.~320--332, 2017.

\bibitem{sekhar2024multi}
V.~N. Sekhar, M.~D. Kumar, S.~K. Tippabhotla, B.~C. Rao, I.~C. Daniel, S.~C. Chong, and V.~S. Rao, ``Multi-chip stacked memory module development using chip to wafer (c2w) hybrid bonding for heterogeneous integration applications,'' in {\em 2024 IEEE 74th Electronic Components and Technology Conference (ECTC)}, IEEE, 2024.

\bibitem{micronddr4}
{Micron Technology}, ``{Micron DDR4 Verilog Model},'' 2018.

\bibitem{stc_report}
``{Nvidia a100 tensor core gpu architecture}.'' \url{https://images.nvidia.com/aem-dam/en-zz/Solutions/data-center/nvidia-ampere-architecture-whitepaper.pdf}.

\bibitem{wu2022sparseloop}
Y.~N. Wu, P.-A. Tsai, A.~Parashar, V.~Sze, and J.~S. Emer, ``Sparseloop: An analytical approach to sparse tensor accelerator modeling,'' in {\em 2022 55th IEEE/ACM International Symposium on Microarchitecture (MICRO)}, pp.~1377--1395, IEEE, 2022.

\bibitem{orenes2024muchisim}
M.~Orenes-Vera, E.~Tureci, M.~Martonosi, and D.~Wentzlaff, ``Muchisim: A simulation framework for design exploration of multi-chip manycore systems,'' in {\em 2024 IEEE International Symposium on Performance Analysis of Systems and Software (ISPASS)}, pp.~48--60, IEEE, 2024.

\end{thebibliography}

\clearpage

\end{document}